\documentclass[%
 reprint,
superscriptaddress,
 amsmath,amssymb,
 floatfix,
 aps,
 prl,
]{revtex4-1}
\usepackage{graphicx}

\usepackage{dcolumn}
\usepackage{bm}

\usepackage{color}
\usepackage{braket}
\usepackage[normalem]{ulem}
\usepackage{bbm}

\newcommand{\Tr}{\text{Tr}}

\newcommand{\E}[2]{\mathbbm{E}_{#1}\left[#2\right]}
\usepackage[linkcolor={blue},citecolor={blue},colorlinks=true]
{hyperref}
\begin{document}
\title{Entanglement Entropy of Non-Hermitian Eigenstates and the Ginibre Ensemble}

\author{Giorgio Cipolloni}
\email{gc4233@princeton.edu}
 \affiliation{Princeton Center for Theoretical Science, Princeton University, Princeton, NJ 08544, USA}
\author{Jonah Kudler-Flam}%
 \email{jkudlerflam@ias.edu}
  \affiliation{Princeton Center for Theoretical Science, Princeton University, Princeton, NJ 08544, USA}
  \affiliation{School of Natural Sciences, Institute for Advanced Study, Princeton, NJ 08540 USA}
\affiliation{Kadanoff Center for Theoretical Physics, University of Chicago, Chicago, IL 60637, USA}


\begin{abstract}
Entanglement entropy is a powerful tool in characterizing universal features in quantum many-body systems. In quantum chaotic Hermitian systems, typical eigenstates have near maximal entanglement with very small fluctuations. Here, we show that for Hamiltonians displaying non-Hermitian many-body quantum chaos, modeled by the Ginibre ensemble, the entanglement entropy of typical eigenstates is greatly suppressed. The entropy does not grow with the Hilbert space dimension for sufficiently large systems and the fluctuations are of equal order. We derive the novel entanglement spectrum that has infinite support in the complex plane and strong energy dependence. We provide evidence of universality and similar behavior is found in the non-Hermitian Sachdev-Ye-Kitaev (nSYK) model, indicating the general applicability of the Ginibre ensemble to dissipative many-body quantum chaos.
\end{abstract}

\maketitle

\textit{Introduction.}--- 
In quantum mechanical systems, it is standard to take the Hamiltonian of the system to be Hermitian, ensuring the reality of the energy spectrum. However, relaxing this Hermiticity has proven to lead to many novel and unexpected phenomena \cite{2020AdPhy..69..249A}. These phenomena are not mere theoretical curiosities, but physically relevant, describing the physics of open quantum systems.

Given the widespread applicability of non-Hermitian physics, it is natural to ask what features are universal. This is our impetus for combining two unifying subjects in the context of non-Hermitian many-body physics, entanglement and random matrix theory. Entanglement entropy has been an indispensable tool characterizing many-body physics, with milestone results in gapped \cite{2007JSMTE..08...24H}, critical \cite{2003PhRvL..90v7902V}, topological \cite{2006PhRvL..96k0405L, 2006PhRvL..96k0404K}, holographic \cite{2006PhRvL..96r1602R}, and dynamical \cite{2005JSMTE..04..010C} systems. Entanglement theory has only recently been applied to non-Hermitian physics, with some of the main achievements coming from characterizations of non-unitary conformal field theories \cite{2015JPhA...48dFT01B, 2015NuPhB.896..835B, 2016JPhA...49o4005B,2017PhRvL.119d0601C,2018ScPP....4...31D,2020PhRvR...2c3069C,2021arXiv210713006T}.

The goal of this Letter is to move away from these ground state studies to the generic properties of \textit{typical eigenstates}. This is of particular interest in the context of the emerging field of dissipative quantum many-body chaos \cite{2019PhRvL.123i0603H,2019PhRvL.123y4101A,2020PhRvX..10b1019S, 2021PhRvL.127q0602L,2021arXiv211212109S,2021arXiv211213489K,2019PhRvL.123n0403D,2020PhRvL.124j0604W,2021PhRvR...3b3190S}. Our strategy is to analyze the eigenstates of the complex Ginibre ensemble, whose matrix elements are independent and identically distributed (i.i.d.) complex Gaussian random variables \cite{1965JMP.....6..440G}. This is our proposed analogy to the typical eigenstates frequently used in Hermitian systems that are eigenstates of the Gaussian Unitary Ensemble (equivalently ``Haar random'' states) \cite{1993PhRvL..71.1291P}. The Ginibre ensemble has been demonstrated to universally emerge in non-Hermitian many-body quantum chaotic systems \cite{2019PhRvL.123y4101A,2020PhRvX..10b1019S, 2019PhRvL.123i0603H,2021PhRvL.127q0602L, 2022PhRvX..12b1040G,Chan_forthcoming}. This is anticipated by the dissipative analog of the Berry-Tabor and Bohigas-Giannoni-Schmit conjectures \cite{PhysRevLett.61.1899,PhysRevLett.62.2893}.

Non-Hermitian Hamiltonians, $H$, have distinct left and right eigenvectors, $\ket{L_i}$ and $\ket{R_i}$, residing in an $N$-dimensional Hilbert space, that form a {biorthonormal} basis $\bra{L_i}R_j\rangle = \delta_{ij}.$
Following the biorthogonal formulation of quantum mechanics \cite{2014JPhA...47c5305B}, we choose the density matrix of an eigenstate to inherit the non-Hermiticity of the Hamiltonian
\begin{align}
\label{eq:denmat}
  \rho^{(i)} := \ket{R_i}\bra{L_i}.
\end{align}
With this choice, the Heisenberg evolution of general density matrices remains $i \partial_t \rho= [H,\rho] $.

We consider a bipartition of the Hilbert space $\mathcal{H} = \mathcal{H}_A \otimes \mathcal{H}_B$ with sub-Hilbert space dimensions $N_A$ and $N_B$. Performing the partial trace on $\mathcal{H}_B$, we arrive at the reduced density matrix
\begin{align}
  \rho_A^{(i)} := \Tr_B \rho^{(i)} .
  \label{rhoA_nonherm}
\end{align}
This describes the state localized to subsystem $A$ because the expectation values of all observables are captured by the reduced state $\langle \mathcal{O}_A\rangle =\Tr \left(\rho^{(i)}_A \mathcal{O}_A\right) $.

While \eqref{rhoA_nonherm} still has unit trace, its eigenvalues are generally complex. To accommodate, we use a generalized definition of the entanglement entropy \cite{2021arXiv210713006T}\footnote{$-\Tr \rho \log \rho$ becomes ambiguous due to the complex arguments of the logarithm. Choosing the principal value, the answer becomes $O(N_A)$ which we do not expect to be useful (see Supplemental Material, which includes \cite{1962JMP.....3.1191D, 2009arXiv0907.5605E, bourgade2018random,2013arXiv1312.1301B, 2021arXiv210306730C, 2020arXiv200508425M, 2022arXiv22XXXXXXXC2}). }
\begin{align}
  S_{vN}\left(\rho_A\right) = -\Tr \rho_A \log \left|\rho_A\right|,
\end{align}
which reduces to the standard entanglement entropy for the Hermitian case. While its quantum information theoretic interpretation is not yet entirely understood, it obeys desirable properties such as $S_{vN} \neq 0$ only if $A$ and $B$ are entangled, it is amenable to path integral constructions, and has been useful in characterizing non-unitary conformal field theories \cite{2017PhRvL.119d0601C,2018ScPP....4...31D,2020PhRvR...2c3069C,2021arXiv210713006T}.

The entropy as a function of $\log N_A$ is referred to as the ``Page curve'' \cite{1993PhRvL..71.1291P} and has been the topic of intense study in both many-body and quantum gravitational physics \cite{1993PhRvL..71.3743P, 2017PhRvL.119v0603V, 2020JHEP...09..002P,2019JHEP...12..063A}. In this Letter, we compute the Page curve for non-Hermitian systems by first identifying the structure of typical reduced density matrices, evaluating {their} eigenvalues (called the \textit{entanglement spectrum}), then computing the expectation of the entanglement entropy and its variance. We numerically demonstrate that our results exhibit \textit{universality} by studying other random matrix ensembles as well as the nSYK model.

\textit{Structure of Reduced Density Matrix.}---To get oriented, we review the Hermitian case where one considers eigenvectors of the Gaussian Unitary Ensemble (GUE). The eigenvectors on $\mathcal{H}_A\otimes \mathcal{H}_B$ can be written as
\begin{align}
  \ket{\Psi} = \sum_{i=1}^{N_A} \sum_{\alpha=1}^{N_B} X_{i\alpha}\ket{i}_A \otimes \ket{\alpha}_B,
\end{align}
where the states in the sum are orthonormal bases for the sub-Hilbert spaces and the $X_{i\alpha}$'s (matrix elements of $X$) are i.i.d.~complex Gaussian random variables with variance $N^{-1}$. The random induced states on $\mathcal{H}_A$ are $\rho_A = XX^{\dagger}$,
defining the celebrated Wishart ensemble \cite{wishart1928generalised}. The spectrum of Wishart matrices is given by the Marchenko–Pastur distribution. The entropy is consequently evaluated to
\begin{align}
  \E{X}{S_{vN}} = \log N_A -\frac{N_A}{2N_B}, \quad N_A \leq N_B.
  \label{page_eq}
\end{align}
The entropy is extremely close to the upper bound of $\log N_A$. This scales \textit{extensively} with the system size and is independent of the eigenvalue {location}, which we will soon see is not the case for non-Hermitian systems.

We seek the non-Hermitian analog of $XX^{\dagger}$. To describe the structure of $\rho_A^{(i)}$, we use the so-called ``Hermitization trick'' \cite{1997NuPhB.504..579F}. Define
\begin{equation}
\label{eq:hemr}
H^z:=\left(\begin{matrix}
0 & H-z \\
(H-z)^{\dagger} & 0
\end{matrix}\right),
\end{equation}
with $z\in \mathbb{C}$ and $H$ is drawn from the Ginibre ensemble. We point out that
\begin{equation}
\label{eq:hermnonherm}
\lambda\in \mathrm{Spec}(H)\Leftrightarrow 0\in \mathrm{Spec}(H^\lambda).
\end{equation}
This is the key observation that will enable us to compute the spectrum of $\rho_A^{(i)}$ and consequently its entaglement entropy. We denote the eigenvalues of $H^z$ by $E_{\pm i}^z$ and by $\ket{{\bf w}_{\pm i}^z}$ the corresponding orthogonal eigenvectors. The {chiral symmetry} of $H^z$ induces a symmetric spectrum around zero, i.e. $E_i^z\ge 0$ and $E_{-i}^z=-E_i^z$; accordingly the eigenvectors $\ket{{\bf w}_{\pm i}^z}$ are of the form $\ket{{\bf w}_{\pm i}^z}=(\ket{{\bf u}_i^z},\pm\ket{{\bf v}_i^z})$, with $\ket{{\bf u}_i^z},\ket{{\bf v}_i^z}\in\mathbb{C}^N$. $E_i^z$ deterministically coincides with the singular values of $H-z$, and $\ket{{\bf u}_i^z},\ket{{\bf v}_i^z}$ denote the corresponding left and right singular vectors, i.e.
\begin{equation}
\label{eq:defsingv}
(H-z)\ket{{\bf v}_i^z}=E_i^z\ket{{\bf u}_i^z}, \quad
(H-z)^{\dagger}\ket{{\bf u}_i^z}=E_i^z\ket{{\bf v}_i^z}.
\end{equation}
The representations \eqref{eq:hemr} and \eqref{eq:defsingv} are completely equivalent.

Let $\lambda_i$ be the eigenvalue with corresponding right and left eigenvectors $\ket{R_i}$, $\ket{L_i}$ from \eqref{eq:denmat}. Then by \eqref{eq:hermnonherm}--\eqref{eq:defsingv}, it follows that
\begin{equation}
\label{eq:eigvrel}
\ket{{\bf v}_1^{\lambda_i}}=\frac{\ket{L_i}}{\lVert L_i\rVert}, \qquad \ket{{\bf u}_1^{\lambda_i}}=\frac{\ket{R_i}}{\lVert R_i\rVert}.
\end{equation}
We introduce the notations $\ket{{\bf v}}:=\ket{{\bf v}_1^{\lambda_i}}$, $\ket{\bf u}:=\ket{{\bf u}_1^{\lambda_i}}$, and thus find that
\begin{equation}
\label{eq:densA}
\rho_A^{(i)}=\frac{\mathrm{Tr}_B[|{\bf u}\rangle\langle {\bf v}|]}{\bra{{\bf u}} {\bf v}\rangle}.
\end{equation}
The key point is that for fixed deterministic $z$ we can compute the distribution of $\ket{{\bf u}_i^z}, \ket{{\bf v}_i^z}$ using Hermitian techniques such as the Dyson Brownian motion (DBM) for eigenvectors introduced in \cite{2013arXiv1312.1301B} (see also \cite{2020arXiv200508425M,2022CMaPh.391..401B,2021arXiv210306730C,2022arXiv220301861C, bourgade2018random}). We explain this in more detail in the supplemental material. This would not have been possible analyzing the non-Hermitian eigenvectors $R_i$ and $L_i$ directly since there is no known non-Hermitian analogue for eigenvector DBM. {By \eqref{eq:eigvrel}, we need to study the case $z=\lambda_i$, i.e. when $z$ is random}, however we expect (and numerically confirm) {the same distribution as $\ket{{\bf u}_i^z}, \ket{{\bf v}_i^z}$ for fixed $z$.} We remark that for $H$ Ginibre, a similar result can be obtained via Weingarten calculus \cite{2002math.ph...5010C}, however we decided to rely on DBM since this approach applies to more general ensembles for $H$ (see \emph{Universality} below). 

We will now study the spectrum of $\rho_A^{(i)}$ conditioned on the event that $\lambda_i=z$, for some $|z|<1$. In order to keep the argument simple and concise we neglect the case $|z|\approx 1$; the analysis in this regime would be analogous except for the fact that the distribution of $(\lVert L_i\rVert\lVert R_i\rVert)^{-1}$ would be more complicated compared to what we have below \eqref{eq:gamma} (see \cite{2018CMaPh.363..579F}).

Since $H$ is a Ginibre matrix, the singular vectors $\ket{{\bf v}_i}$ of $H-z$ are Haar unitary distributed {(here $\ket{{\bf v}_1}=\ket{{\bf v}}$)}.\footnote{Note that if we had taken the density matrix to be Hermitian i.e.~$\ket{R}\bra{R}$ or $\ket{L}\bra{L}$, this would imply that the reduced density matrices are, once again, drawn from the Wishart ensemble.} We now write $\ket{\bf u}$ in the $\ket{{\bf v}_i}$ basis:
\begin{align}
  \ket{\textbf{u}} = \sum_{i} c_i \ket{\textbf{v}_i}, \qquad c_i:=\bra{{\bf v}_i} {\bf u}\rangle.
\end{align}
The coefficient $c_1$ is distributed as 
\begin{align}
\label{eq:gamma}
 c_1 = \braket{\textbf{u}|\textbf{v}_1} = \sqrt{ \frac{\gamma_2}{N(1-|z|^2)}}.
\end{align}
as computed in \cite{2018CMaPh.363..579F, 2018arXiv180101219B}. In particular, for the distribution of $c_1$ no DBM is required. $\gamma_2$ is a random variable drawn from the Gamma distribution with shape parameter $2$, i.e. its density is given by~$xe^{-x}$. The rest of the $c_i$'s, for $i\ge 2$, are i.i.d.~standard complex Gaussian random variables and independent of $c_1$. More precisely, using DBM \cite{2013arXiv1312.1301B, 2020arXiv200508425M,2022CMaPh.391..401B,2021arXiv210306730C,2022arXiv220301861C, bourgade2018random} we can show that any finite (independent of $N$) collection of $c_i$'s, for $i\ge 2$, converge to i.i.d.~standard complex Gaussians. For Ginibre one can expect, for instance using Weingarten calculus, that this convergence holds for all the $c_i$'s with $i\ge 2$. This is also confirmed numerically below. {We point out that to use DBM we write} \begin{equation}
\label{eq:mainnonherm}
c_i:=\bra{{\bf v}_i}{\bf u}\rangle=\bra{{\bf w}_1^z} F \ket{{\bf w}_i^z}, \qquad F:=\left(\begin{matrix} 0 & 0 \\ 1 & 0 \end{matrix}\right),
\end{equation} with $\ket{{\bf w}_i^z}$ being defined below \eqref{eq:hermnonherm}, since eigenvector overlaps of this form are well understood for Hermitian matrices using DBM  \cite{2013arXiv1312.1301B,2020arXiv200508425M,2022CMaPh.391..401B,2021arXiv210306730C,2022arXiv220301861C, bourgade2018random}. See the supplemental material for a gentle explanation of this approach.

We thus find that
\begin{align}
  \rho_{AB} = \sum_{i}\frac{c_i}{c_1}\ket{\textbf{v}_i}\bra{\textbf{v}_1},
  \label{rhoABfull}
\end{align}
which has unit trace but is non-Hermitian, with the $c_i$'s distributed as described above. In the $\mathcal{H}_A\otimes \mathcal{H}_B$ basis, we have
\begin{align}
\label{eq:vvv}
  \ket{\textbf{v}_i} = \sum_{j\alpha} X^{(i)}_{j\alpha}\ket{j}_A \ket{\alpha}_B ,
\end{align}
where the $X^{(i)}$'s are $N_A \times N_B$ rectangular matrices with i.i.d.~Gaussian variables with variance $N^{-1}$ because {the singular vectors, $\ket{\textbf{v}_i}$,} are Haar distributed. We neglect the normalization because it concentrates around one at large-$N$. {The singular vectors} are correlated with each other only in that they are orthonormal, an unimpactful subtlety that we ignore in the limits we consider. The reduced density matrix is thus given by
\begin{align}
\label{eq:reduced}
  \rho_{A} &= \sum_{i}\frac{c_i}{c_1} X^{(i)}X^{(1)\dagger} ,
\end{align}
defining a non-Hermitian analog of the Wishart ensemble. Distinct non-Hermitian analogs of the Wishart ensemble have been studied in the math literature \cite{2011arXiv1104.5203A,2010PhRvE..82f1114B, 2021AnHP...22.1035A}, {though these ensembles have no clear interpretation as density matrices for quantum systems}. The most striking difference is that we will see that \eqref{eq:reduced} has non-compact support with a heavy tail, unlike the compactly supported eigenvalue spectra previously studied{. This is a consequence of the correlation of the left and right eigenvectors encoded in $c_1$, which the models in \cite{2011arXiv1104.5203A,2010PhRvE..82f1114B, 2021AnHP...22.1035A} are not able to capture.} This makes the analysis much more delicate in the current case. As shown in \cite{2010PhRvE..82f1114B}, the product of rectangular i.i.d.~matrices gives a matrix with the spectrum of a Ginibre matrix for $1\ll N_A \ll N_B$ rescaled by $N^{-1/2}$. When $i = 1$, this limit leads to the normalized identity matrix plus a matrix with the spectrum of a GUE matrix suppressed by $N^{-1/2}$ and hence irrelevant. In total, the density matrix takes the form 
\begin{align}
  \rho_A = \frac{\mathbbm{1}_A}{N_A} + \frac{M_{GUE}}{\sqrt{N}} +\sqrt{\frac{1-|z|^2}{\gamma_2 N}} \sum_{i =2}^N {c_i} M^{(i)}_{Ginibre}.
\end{align}
By the central limit theorem (CLT), we can add all the random matrices at large $N$ to find 
\begin{align}
  \rho_A = \frac{\mathbbm{1}_A}{N_A} +\sqrt{\frac{1-|z|^2}{\gamma_2}} M,
  \label{rhoA_approx}
\end{align}
where $M$ is a Ginibre matrix independent of $\gamma_2$. Note that to go from \eqref{eq:reduced} to \eqref{rhoA_approx}, we did not need that $X^{(i)}X^{(1)\dagger}$ is approximately Ginibre and that the $c_i$'s for $i\ge 2$ are i.i.d.; we only needed to ensure that the CLT for the entries of $\rho_A$ held. This remark will be relevant for the \emph{Universality} discussion below when matrices $H$ with not necessarily Gaussian entries are considered. We thus find that reduced density matrices of the Ginibre ensemble are Ginibre themselves, with a random, eigenvalue dependent, scaling and deterministic shift.

\begin{figure}
  \centering
  \includegraphics[width = .4\textwidth]{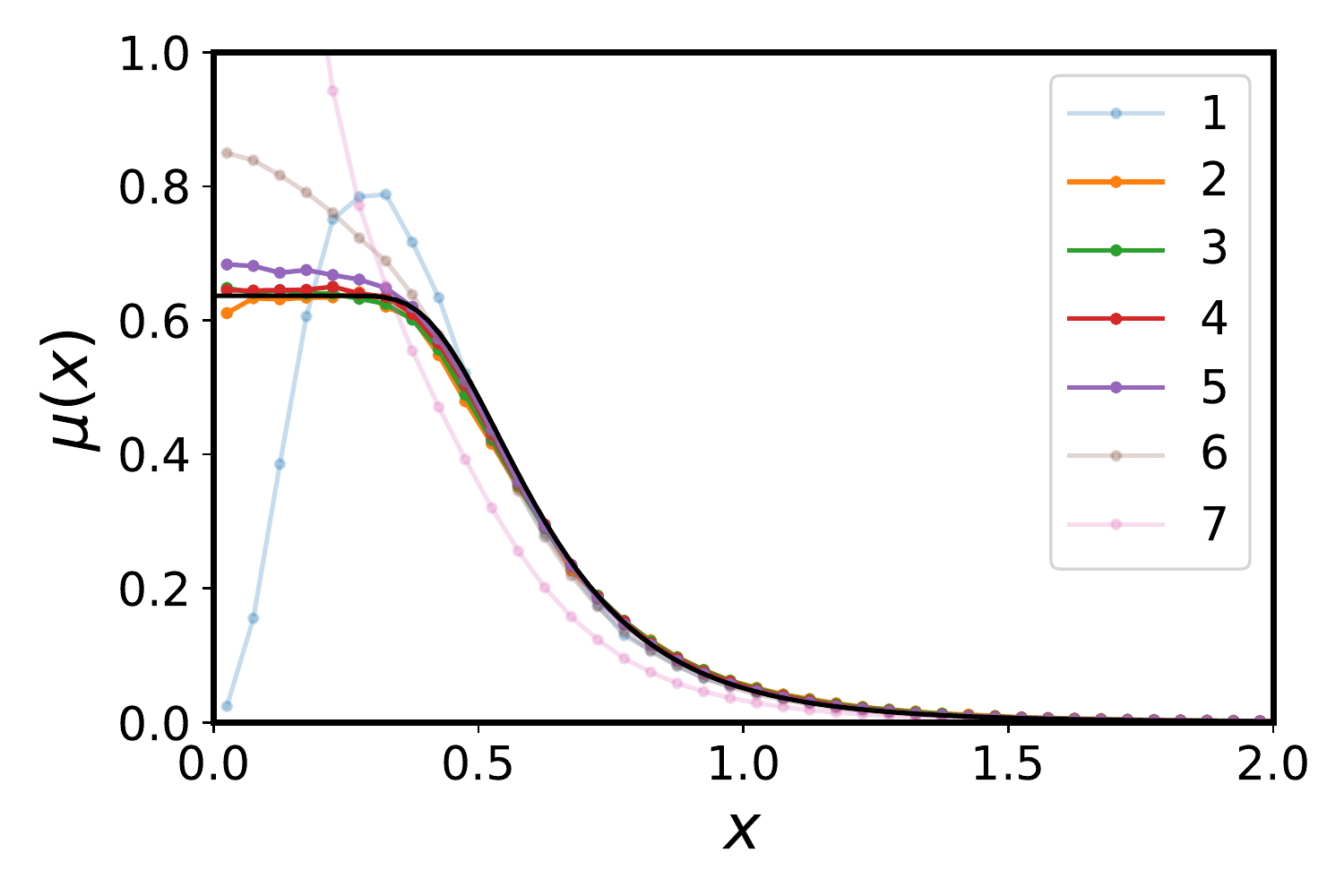}
  \caption{The entanglement spectrum is shown for various values of $\log_2 N_A$ (labeled in the legend) and $N = 2^{14}$. The data is averaged over all eigenstates and compared with the large-$N$ formula, \eqref{ansatz_spectrum}, displayed as the solid black line. There are clear deviations from \eqref{ansatz_spectrum} outside of the $1 \ll N_A \ll N_B$ regime.}
  \label{finite_N_spectrum_fig}
\end{figure}

\textit{Entanglement Spectrum.}---Now that we understand the structure of the non-Hermitian ensemble defining the reduced density matrix, we compute the entanglement spectrum. Famously, the spectrum of Ginibre matrices is uniformly distributed on the unit circle \cite{1965JMP.....6..440G}. In the $1 \ll N_A \ll N_B$ regime, the entanglement spectrum, conditioned on $\gamma_2$, is therefore given by a shifted circular law due to the Ginibre matrix in \eqref{rhoA_approx}
\begin{align}
  \mu(x, \gamma_2) =
  \begin{cases}
  \frac{\gamma_2}{\pi }, & x < \gamma_2^{-1/2}
  \\
  0 , & x > \gamma_2^{-1/2}
  \end{cases},
\end{align}
where $x := \frac{\left| N_A^{-1}-\lambda \right|}{\sqrt{1-|z|^2}}$.
Integrating this distribution over $\gamma_2$,
\begin{align}
    \mu(x) = \frac{1}{\pi} \int_0^{x^{-2}} d\gamma_2 \gamma_2^2e^{-\gamma_2},
\end{align}
we then find
\begin{align}
  \label{ansatz_spectrum}
  \mu(x) = \frac{1}{\pi}\left(2-e^{-\frac{1}{x^2}} \left(\frac{1+2
  x^2+2x^4}{ x ^4}\right)\right).
\end{align}
Note that $\mu(x)$ is rotationally invariant.
For $N_A > N_B$, the spectrum is identical with $N_A \leftrightarrow N_B$ and the addition of $N_A-N_B$ eigenvalues at $\lambda = 0$.
This spectrum has infinite support with a very heavy tail, decaying only as $|\lambda|^{-6}$ at large $|\lambda|$, in stark contrast with the compactly supported eigenvalue spectra of \cite{2011arXiv1104.5203A,2010PhRvE..82f1114B, 2021AnHP...22.1035A}. In Figure \ref{finite_N_spectrum_fig}, we show the very good agreement between \eqref{ansatz_spectrum} and numerical data for small matrices. We lack an analytical expression for the spectrum at $N_A = N_B$, though numerically show the accuracy of \eqref{rhoABfull} in all regimes in the supplemental material.

\begin{figure}
  \centering
  \includegraphics[width = .355\textwidth]{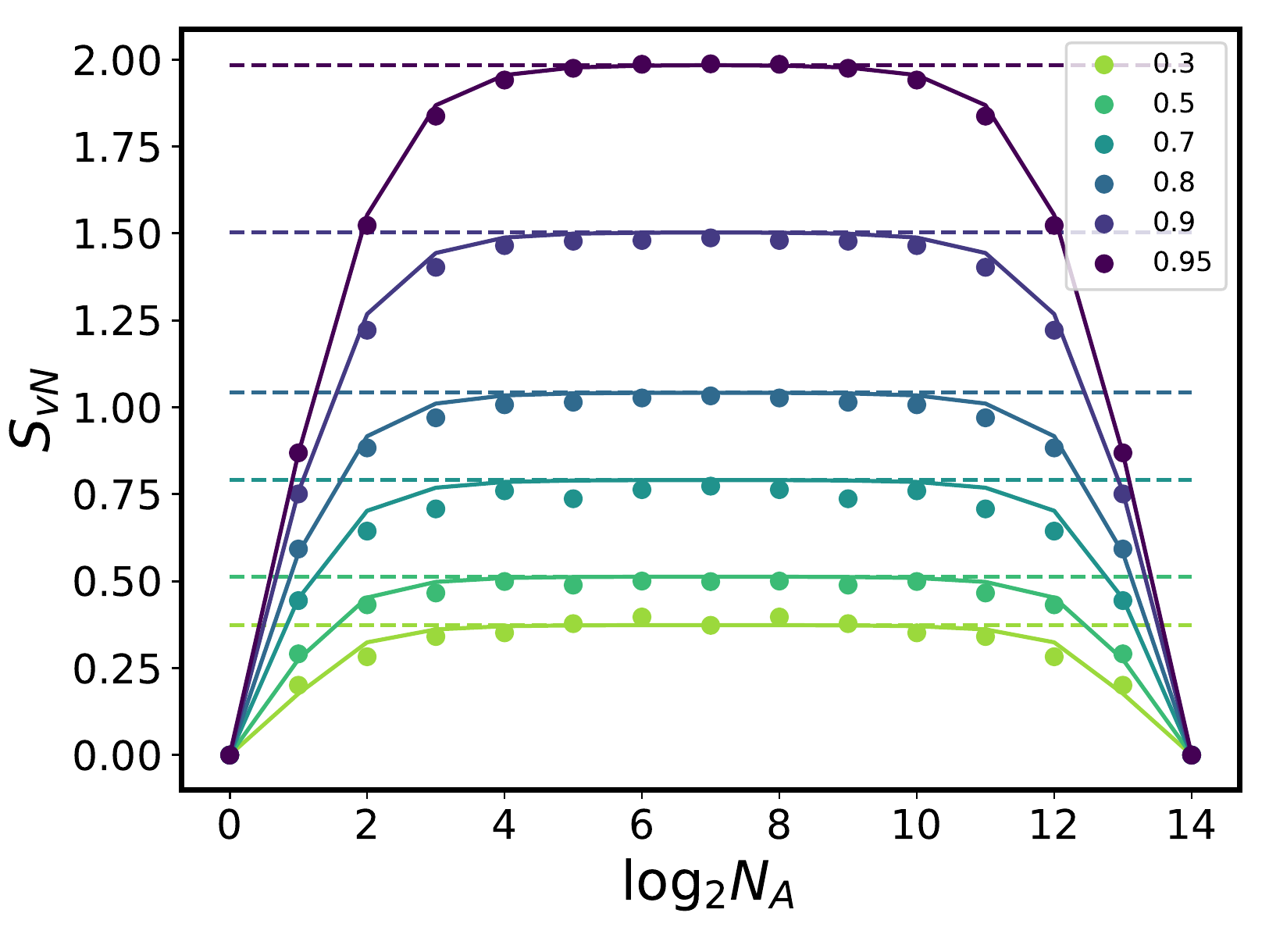}
  \caption{The von Neumann entropy (base $2$) for eigenvectors of Ginibre matrices for various values of $|z|$. The circles are numerical data, the solid line is evaluated from the finite $N$ spectrum \eqref{ansatz_spectrum}, and the dashed line is the asymptotic result \eqref{asymptotic_entropy}. We have taken $N = 2^{14}$ and $10^3$ disorder realizations. Only the real part is plotted because the imaginary part averages to zero.}
  \label{page_nonherm}
\end{figure}

\textit{Page Curve.}---The average entropy is given by
\begin{align}
  \E{M,\gamma_2}{S_{vN}(|z|)} = -N_A\int d\lambda \mu(\lambda) \lambda \log |\lambda| .
\end{align}
Conditioned on $R :=\sqrt{ \frac{1-|z|^2}{\gamma_2}}$, the entropy is
\begin{align}
  \E{M}{S_{vN}(R)} = -\frac{N_A}{\pi R^2}\int_0^{2\pi} d\theta\int_0^R dr r \left(N_A^{-1}+ r e^{i\theta} \right)
  \nonumber  \\
  \times \log \left(N_A^{-2}+\frac{2 r \cos (\theta )}{N_A}+r^2\right).
\end{align}
Expanding in $N_A$, only the $O(N_{A}^0)$ term contributes due to the integral over $\theta$, leading to
\begin{align}
  \E{M}{S_{vN}(R)} = - \log R.
\end{align}
Integrating over $\gamma_2$, we arrive at
\begin{align}
  \label{asymptotic_entropy}
  \E{M,\gamma_2}{S_{vN}(|z|)} = \frac{1-\gamma-\log \left(1-|z|^2\right)}{2},
\end{align}
where $\gamma$ is the Euler–Mascheroni constant. Surprisingly, there is no scaling with $N_A$, in stark contrast to the Hermitian case \eqref{page_eq}. We show remarkable agreement between \eqref{asymptotic_entropy} and small matrices, even at $N_A = N_B$, in Figure \ref{page_nonherm}.

As a consequence of the large fluctuations in the structure of the density matrix, there are large fluctuations in the von Neumann entropy. Therefore, we would like to understand its variance. To do so, we split the variance into three terms
\begin{align}
  \E{M, \gamma_2}{\left|S_{vN}- \E{M, \gamma_2} {S_{vN}}\right|^2} =\E{\gamma_2}{\left|\E{M}{ S_{vN}}\right|^2}
  \nonumber
  \\
  - \left|\E{M, \gamma_2} {S_{vN}}\right|^2+\E{M, \gamma_2}{\left|S_{vN}- \E{M}{S_{vN}}\right|^2}. \label{eq:var1}
\end{align}
The first term on the second line is simply the square of \eqref{asymptotic_entropy}. The term in the first line may be analogously computed at large $N_A$ by Taylor expanding the logarithm
\begin{align}
   \E{\gamma_2}{\left|\E{M}{ S_{vN}}\right|^2}= \frac{\pi^2-6}{24}+\left|\E{M, \gamma_2} {S_{vN}}\right|^2.
\end{align}
The most involved term is the final one. Fortunately, the variance of functions of eigenvalues, $f(\lambda_i)$, of Ginibre matrices was analyzed in \cite{2006math......6663R}. There, it was found that
\begin{align}
  \E{M}{\left|\sum_if(\lambda_i) - \E{M}{\sum_if(\lambda_i)} \right|^2} =
  \nonumber
  \\
  \frac{1}{4\pi}\int_{\textbf{D}}d^2z \left|\nabla f\right|^2 + \frac{1}{2}\sum_{k \in \mathbb{Z}}|k|\left|\hat{f}(k)\right|^2, \label{eq:var2}
\end{align}
where $\textbf{D}$ is the unit disk and $\hat{f}(k)$ is the $k^{th}$ Fourier mode of $f$ on the perimeter of the disk. For the entropy, we must take
\begin{align}
  f(z) = -\left( N_A^{-1} + R z \right)\log \left|N_A^{-1} + R z \right|.
\end{align}
After averaging over $\gamma_2$, the three terms are $N_A$-independent at large $N_A$, thus the same order as the mean \eqref{asymptotic_entropy}. Interestingly, the variance is monotonically decreasing with $|z|$, ranging between $\sim 0.906$ at $z = 0$ and $\sim 0.161$ at $z = 1$.

\begin{figure}
    \centering
    \includegraphics[width = .48\textwidth]{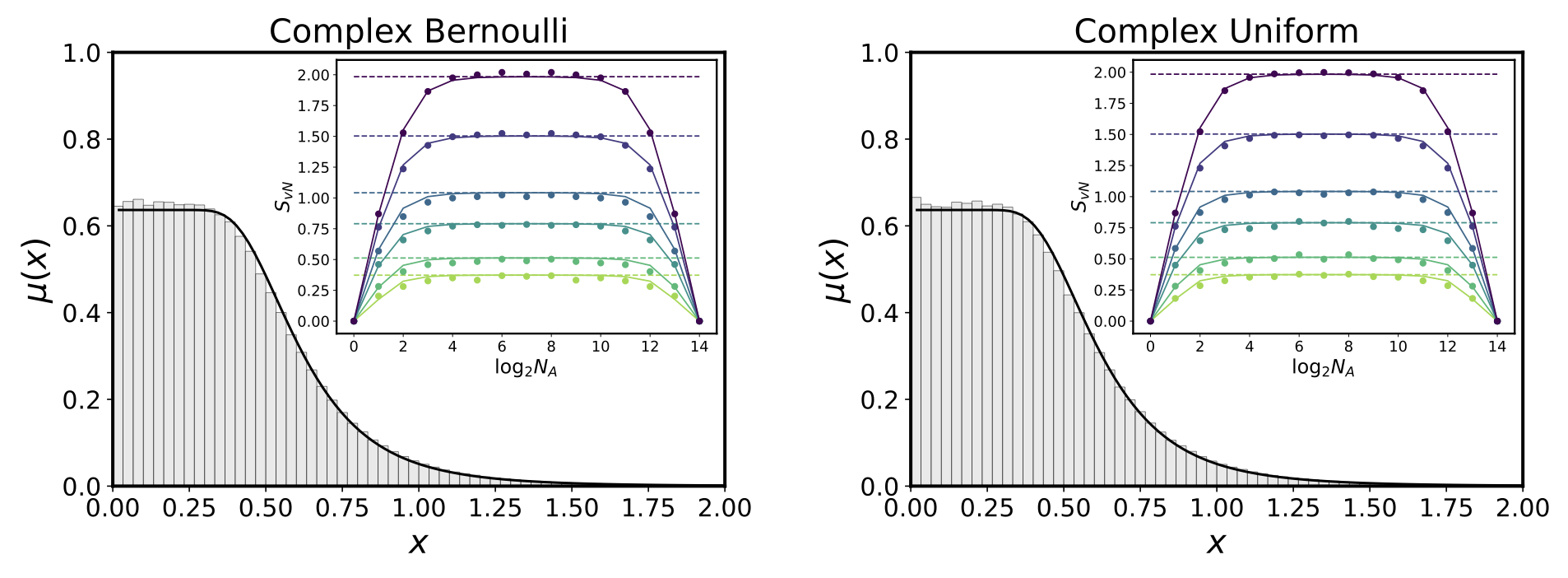}
    \caption{The entanglement spectrum (main plot) and Page curve (upper right) for the eigenvectors of random complex Bernoulli matrices (entries are independent $\pm$ real and imaginary parts) and random complex uniform matrices (entries are uniformly drawn from the complex unit circle). For the spectrum, we take $N_A = 2^4$ and $N = 2^{14}$ to ensure $1 \ll N_A \ll N_B$. The black line is the analytic result for Ginibre matrices \eqref{ansatz_spectrum}. The Page curves precisely agree with the Ginibre Page curve from Figure \ref{page_nonherm}.}
    \label{page_Bernoulli}
\end{figure}

\textit{Universality.}---It is clearly important to understand if our results exhibit \textit{universality}. A similar analysis to the one performed above holds for left and right eigenvectors of more general non-Hermitian matrix ensembles, i.e. for matrices $H$ with i.i.d.~entries but not necessarily with Gaussian distribution, matrices with independent (but not necessarily identically distributed) entries, and even for matrices with some correlation structure; however the precise limiting constant may differ from  \eqref{asymptotic_entropy}. The common feature in all these models is that we expect to get an $N_A$-independent entropy. This is motivated from the fact that the DBM arguments in \cite{2013arXiv1312.1301B, 2020arXiv200508425M,2022CMaPh.391..401B,2021arXiv210306730C,2022arXiv220301861C, bourgade2018random} and reviewed in the supplemental material hold true for fairly general Hermitian ensembles. In these cases the vectors $\ket{{\bf v}_i^z}$ will not be Haar distributed, but the CLT still holds so the
approximation in \eqref{rhoA_approx} will be valid.
We expect the scaling constant $(1-|z|^2)^{-1/2}$ to be dependent only on the shape of the eigenvalue distribution, but the $\gamma_2$ distribution to be universal. For matrices obeying the circular law, we demonstrate the $(1-|z|^2)^{-1/2}$ scaling to be the correct one in Figure \ref{page_Bernoulli} for two random matrix ensembles that are very different than Ginibre, suggesting universality.

\begin{figure}
  \centering
  \includegraphics[width = .48\textwidth]{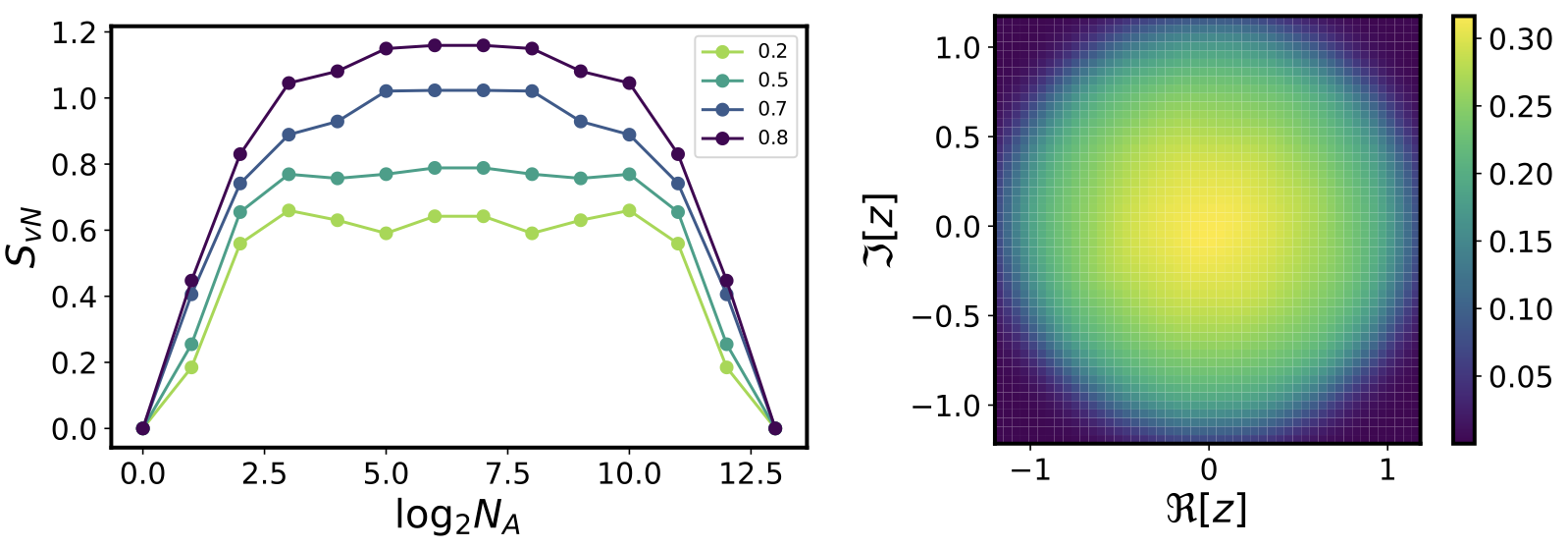}
  \caption{Left: The von Neumann entropy (base 2) of the non-Hermitian SYK model averaged over various values of $|z|$ with $q= 4$ and $N = 26$. Right: The corresponding energy eigenvalue spectrum. The entropy shows a plateau, increasing with $|z|$, akin to the Ginibre ensemble.}
  \label{nSYK_spectrum_N26}
\end{figure}

It is additionally important to consider bona fide Hamiltonian systems such as the nSYK model of $N$ Majorana fermions
\begin{align}
  H_{nSYK} = \sum^N_{i_1 < i_2 <\dots<i_q}\left(J_{i_1i_2\dots i_q}+iM_{i_1i_2\dots i_q} \right)\psi_{i_1}\psi_{i_2}\dots \psi_{i_q},
\end{align}
where $J_{i_1i_2\dots i_q}$ and $M_{i_1i_2\dots i_q} $ are i.i.d.~real Gaussian random variables with zero mean, variance $\frac{2}{N^{q-1}}$, and $\{\psi_i, \psi_j \} = 2\delta_{ij}$. 
For even $q$ and $N \mod 8$ is $2$ or $6$, the Hamiltonian is in the complex Ginibre symmetry class \cite{2020PhRvR...2b3286H, 2022PhRvX..12b1040G}. We show the rotationally symmetric but non-uniform eigenvalue distribution and average entropy in Figure \ref{nSYK_spectrum_N26}. ``Unfolding'' the eigenvalue non-uniformity warrants further attention to compare quantitatively with the Ginibre ensemble.

\textit{Discussion.}---In this Letter, we have presented the entanglement spectrum and entanglement entropy of eigenvectors of Ginibre matrices, relying on the determination of the novel structure of correlations in the density matrix. We found that the entanglement spectrum is non-compactly supported, with eigenvalue densities decaying at infinity with a heavy $|\lambda|^{-6}$ tail in the complex plane. This led us to find a Page curve that did not scale with the system size, vastly suppressed as compared to the Hermitian Page curve. Moreover, we found the Page curve to not be self-averaging with fluctuations of the same order as the mean, in stark contrast with the Hermitian case.

There are many interesting research directions motivated from this work. An important characterization of many-body chaos beyond the entanglement entropy is the eigenstate thermalization hypothesis (ETH) which, motivated by random matrix theory, describes the universal behavior of expectation values of simple observables and their fluctuations \cite{PhysRevA.43.2046, 1994PhRvE..50..888S} (see also \cite{2022CMaPh.391..401B, cipolloni2021eigenstate, bourgade2018random}). In \cite{2022arXiv22XXXXXXXC}, we generalize the ETH to non-Hermitian systems by employing the Ginibre ensemble. Along with providing the compelling prediction that observables have large inter-eigenstate fluctuations (hence no thermalization), this leads to an alternate derivation of \eqref{rhoA_approx}.

Furthermore, it may be interesting to explore the entanglement entropy in different classes of non-Hermitian many-body systems, such as those with symmetries \cite{2020PhRvR...2b3286H,2022PhRvX..12b1040G,2021PhRvA.103f2416M}, those with localization transitions \cite{2019PhRvL.123i0603H}, non-interacting fermions \cite{2022PhRvX..12b1040G}, Liouvillians \cite{2019PhRvL.123n0403D, 2019JPhA...52V5302C, 2019PhRvL.123w4103C,2020JPhA...53D5303S,2020PhRvL.124j0604W,2021Chaos..31b3101L,2021PhRvE.104c4118T}, non-equilibrium systems \cite{2022arXiv220109895T,2022arXiv220605384K}, or many-body scars \cite{2021PhRvR...3d3156P}.
We hope to report on some of these directions in the near future.

\begin{acknowledgments}
\textit{Acknowledgments.}---We would like to thank Amos Chan and Shinsei Ryu for useful discussions and comments. We especially thank Kohei Kawabata for his comments that significantly improved the manuscript. JKF is supported through a Simons Investigator Award to Shinsei Ryu from the Simons Foundation (Award Number: 566166) and by the Institute for Advanced Study and the National Science Foundation under Grant No. PHY-2207584. We use \textsc{QuSpin} for simulating the nSYK model \cite{2017ScPP....2....3W,2019ScPP....7...20W} and thank Laimei Nie for her help in implementation. 
\end{acknowledgments}

\bibliographystyle{apsrev}
\bibliography{main}

\begin{thebibliography}{70}
\expandafter\ifx\csname natexlab\endcsname\relax\def\natexlab#1{#1}\fi
\expandafter\ifx\csname bibnamefont\endcsname\relax
  \def\bibnamefont#1{#1}\fi
\expandafter\ifx\csname bibfnamefont\endcsname\relax
  \def\bibfnamefont#1{#1}\fi
\expandafter\ifx\csname citenamefont\endcsname\relax
  \def\citenamefont#1{#1}\fi
\expandafter\ifx\csname url\endcsname\relax
  \def\url#1{\texttt{#1}}\fi
\expandafter\ifx\csname urlprefix\endcsname\relax\def\urlprefix{URL }\fi
\providecommand{\bibinfo}[2]{#2}
\providecommand{\eprint}[2][]{\url{#2}}

\bibitem[{\citenamefont{{Ashida} et~al.}(2020)\citenamefont{{Ashida}, {Gong},
  and {Ueda}}}]{2020AdPhy..69..249A}
\bibinfo{author}{\bibfnamefont{Y.}~\bibnamefont{{Ashida}}},
  \bibinfo{author}{\bibfnamefont{Z.}~\bibnamefont{{Gong}}}, \bibnamefont{and}
  \bibinfo{author}{\bibfnamefont{M.}~\bibnamefont{{Ueda}}},
  \bibinfo{journal}{Advances in Physics} \textbf{\bibinfo{volume}{69}},
  \bibinfo{pages}{249} (\bibinfo{year}{2020}), \eprint{2006.01837}.

\bibitem[{\citenamefont{{Hastings}}(2007)}]{2007JSMTE..08...24H}
\bibinfo{author}{\bibfnamefont{M.~B.} \bibnamefont{{Hastings}}},
  \bibinfo{journal}{Journal of Statistical Mechanics: Theory and Experiment}
  \textbf{\bibinfo{volume}{2007}}, \bibinfo{pages}{08024}
  (\bibinfo{year}{2007}), \eprint{0705.2024}.

\bibitem[{\citenamefont{{Vidal} et~al.}(2003)\citenamefont{{Vidal}, {Latorre},
  {Rico}, and {Kitaev}}}]{2003PhRvL..90v7902V}
\bibinfo{author}{\bibfnamefont{G.}~\bibnamefont{{Vidal}}},
  \bibinfo{author}{\bibfnamefont{J.~I.} \bibnamefont{{Latorre}}},
  \bibinfo{author}{\bibfnamefont{E.}~\bibnamefont{{Rico}}}, \bibnamefont{and}
  \bibinfo{author}{\bibfnamefont{A.}~\bibnamefont{{Kitaev}}},
  \bibinfo{journal}{\prl} \textbf{\bibinfo{volume}{90}}, \bibinfo{eid}{227902}
  (\bibinfo{year}{2003}), \eprint{quant-ph/0211074}.

\bibitem[{\citenamefont{{Levin} and {Wen}}(2006)}]{2006PhRvL..96k0405L}
\bibinfo{author}{\bibfnamefont{M.}~\bibnamefont{{Levin}}} \bibnamefont{and}
  \bibinfo{author}{\bibfnamefont{X.-G.} \bibnamefont{{Wen}}},
  \bibinfo{journal}{\prl} \textbf{\bibinfo{volume}{96}}, \bibinfo{eid}{110405}
  (\bibinfo{year}{2006}), \eprint{cond-mat/0510613}.

\bibitem[{\citenamefont{{Kitaev} and {Preskill}}(2006)}]{2006PhRvL..96k0404K}
\bibinfo{author}{\bibfnamefont{A.}~\bibnamefont{{Kitaev}}} \bibnamefont{and}
  \bibinfo{author}{\bibfnamefont{J.}~\bibnamefont{{Preskill}}},
  \bibinfo{journal}{\prl} \textbf{\bibinfo{volume}{96}}, \bibinfo{eid}{110404}
  (\bibinfo{year}{2006}), \eprint{hep-th/0510092}.

\bibitem[{\citenamefont{{Ryu} and {Takayanagi}}(2006)}]{2006PhRvL..96r1602R}
\bibinfo{author}{\bibfnamefont{S.}~\bibnamefont{{Ryu}}} \bibnamefont{and}
  \bibinfo{author}{\bibfnamefont{T.}~\bibnamefont{{Takayanagi}}},
  \bibinfo{journal}{\prl} \textbf{\bibinfo{volume}{96}}, \bibinfo{eid}{181602}
  (\bibinfo{year}{2006}), \eprint{hep-th/0603001}.

\bibitem[{\citenamefont{{Calabrese} and {Cardy}}(2005)}]{2005JSMTE..04..010C}
\bibinfo{author}{\bibfnamefont{P.}~\bibnamefont{{Calabrese}}} \bibnamefont{and}
  \bibinfo{author}{\bibfnamefont{J.}~\bibnamefont{{Cardy}}},
  \bibinfo{journal}{Journal of Statistical Mechanics: Theory and Experiment}
  \textbf{\bibinfo{volume}{2005}}, \bibinfo{pages}{04010}
  (\bibinfo{year}{2005}), \eprint{cond-mat/0503393}.

\bibitem[{\citenamefont{{Bianchini}
  et~al.}(2015{\natexlab{a}})\citenamefont{{Bianchini}, {Castro-Alvaredo},
  {Doyon}, {Levi}, and {Ravanini}}}]{2015JPhA...48dFT01B}
\bibinfo{author}{\bibfnamefont{D.}~\bibnamefont{{Bianchini}}},
  \bibinfo{author}{\bibfnamefont{O.}~\bibnamefont{{Castro-Alvaredo}}},
  \bibinfo{author}{\bibfnamefont{B.}~\bibnamefont{{Doyon}}},
  \bibinfo{author}{\bibfnamefont{E.}~\bibnamefont{{Levi}}}, \bibnamefont{and}
  \bibinfo{author}{\bibfnamefont{F.}~\bibnamefont{{Ravanini}}},
  \bibinfo{journal}{Journal of Physics A Mathematical General}
  \textbf{\bibinfo{volume}{48}}, \bibinfo{eid}{04FT01}
  (\bibinfo{year}{2015}{\natexlab{a}}), \eprint{1405.2804}.

\bibitem[{\citenamefont{{Bianchini}
  et~al.}(2015{\natexlab{b}})\citenamefont{{Bianchini}, {Castro-Alvaredo}, and
  {Doyon}}}]{2015NuPhB.896..835B}
\bibinfo{author}{\bibfnamefont{D.}~\bibnamefont{{Bianchini}}},
  \bibinfo{author}{\bibfnamefont{O.~A.} \bibnamefont{{Castro-Alvaredo}}},
  \bibnamefont{and} \bibinfo{author}{\bibfnamefont{B.}~\bibnamefont{{Doyon}}},
  \bibinfo{journal}{Nuclear Physics B} \textbf{\bibinfo{volume}{896}},
  \bibinfo{pages}{835} (\bibinfo{year}{2015}{\natexlab{b}}),
  \eprint{1502.03275}.

\bibitem[{\citenamefont{{Bianchini} and
  {Ravanini}}(2016)}]{2016JPhA...49o4005B}
\bibinfo{author}{\bibfnamefont{D.}~\bibnamefont{{Bianchini}}} \bibnamefont{and}
  \bibinfo{author}{\bibfnamefont{F.}~\bibnamefont{{Ravanini}}},
  \bibinfo{journal}{Journal of Physics A Mathematical General}
  \textbf{\bibinfo{volume}{49}}, \bibinfo{eid}{154005} (\bibinfo{year}{2016}),
  \eprint{1509.04601}.

\bibitem[{\citenamefont{{Couvreur} et~al.}(2017)\citenamefont{{Couvreur},
  {Jacobsen}, and {Saleur}}}]{2017PhRvL.119d0601C}
\bibinfo{author}{\bibfnamefont{R.}~\bibnamefont{{Couvreur}}},
  \bibinfo{author}{\bibfnamefont{J.~L.} \bibnamefont{{Jacobsen}}},
  \bibnamefont{and} \bibinfo{author}{\bibfnamefont{H.}~\bibnamefont{{Saleur}}},
  \bibinfo{journal}{\prl} \textbf{\bibinfo{volume}{119}}, \bibinfo{eid}{040601}
  (\bibinfo{year}{2017}), \eprint{1611.08506}.

\bibitem[{\citenamefont{{Dupic} et~al.}(2018)\citenamefont{{Dupic}, {Estienne},
  and {Ikhlef}}}]{2018ScPP....4...31D}
\bibinfo{author}{\bibfnamefont{T.}~\bibnamefont{{Dupic}}},
  \bibinfo{author}{\bibfnamefont{B.}~\bibnamefont{{Estienne}}},
  \bibnamefont{and} \bibinfo{author}{\bibfnamefont{Y.}~\bibnamefont{{Ikhlef}}},
  \bibinfo{journal}{SciPost Physics} \textbf{\bibinfo{volume}{4}},
  \bibinfo{eid}{031} (\bibinfo{year}{2018}), \eprint{1709.09270}.

\bibitem[{\citenamefont{{Chang} et~al.}(2020)\citenamefont{{Chang}, {You},
  {Wen}, and {Ryu}}}]{2020PhRvR...2c3069C}
\bibinfo{author}{\bibfnamefont{P.-Y.} \bibnamefont{{Chang}}},
  \bibinfo{author}{\bibfnamefont{J.-S.} \bibnamefont{{You}}},
  \bibinfo{author}{\bibfnamefont{X.}~\bibnamefont{{Wen}}}, \bibnamefont{and}
  \bibinfo{author}{\bibfnamefont{S.}~\bibnamefont{{Ryu}}},
  \bibinfo{journal}{Physical Review Research} \textbf{\bibinfo{volume}{2}},
  \bibinfo{eid}{033069} (\bibinfo{year}{2020}), \eprint{1909.01346}.

\bibitem[{\citenamefont{Tu et~al.}(2022)\citenamefont{Tu, Tzeng, and
  Chang}}]{2021arXiv210713006T}
\bibinfo{author}{\bibfnamefont{Y.-T.} \bibnamefont{Tu}},
  \bibinfo{author}{\bibfnamefont{Y.-C.} \bibnamefont{Tzeng}}, \bibnamefont{and}
  \bibinfo{author}{\bibfnamefont{P.-Y.} \bibnamefont{Chang}},
  \bibinfo{journal}{SciPost Physics} \textbf{\bibinfo{volume}{12}},
  \bibinfo{pages}{194} (\bibinfo{year}{2022}).

\bibitem[{\citenamefont{{Hamazaki} et~al.}(2019)\citenamefont{{Hamazaki},
  {Kawabata}, and {Ueda}}}]{2019PhRvL.123i0603H}
\bibinfo{author}{\bibfnamefont{R.}~\bibnamefont{{Hamazaki}}},
  \bibinfo{author}{\bibfnamefont{K.}~\bibnamefont{{Kawabata}}},
  \bibnamefont{and} \bibinfo{author}{\bibfnamefont{M.}~\bibnamefont{{Ueda}}},
  \bibinfo{journal}{\prl} \textbf{\bibinfo{volume}{123}}, \bibinfo{eid}{090603}
  (\bibinfo{year}{2019}), \eprint{1811.11319}.

\bibitem[{\citenamefont{{Akemann} et~al.}(2019)\citenamefont{{Akemann},
  {Kieburg}, {Mielke}, and {Prosen}}}]{2019PhRvL.123y4101A}
\bibinfo{author}{\bibfnamefont{G.}~\bibnamefont{{Akemann}}},
  \bibinfo{author}{\bibfnamefont{M.}~\bibnamefont{{Kieburg}}},
  \bibinfo{author}{\bibfnamefont{A.}~\bibnamefont{{Mielke}}}, \bibnamefont{and}
  \bibinfo{author}{\bibfnamefont{T.}~\bibnamefont{{Prosen}}},
  \bibinfo{journal}{\prl} \textbf{\bibinfo{volume}{123}}, \bibinfo{eid}{254101}
  (\bibinfo{year}{2019}), \eprint{1910.03520}.

\bibitem[{\citenamefont{{S{\'a}}
  et~al.}(2020{\natexlab{a}})\citenamefont{{S{\'a}}, {Ribeiro}, and
  {Prosen}}}]{2020PhRvX..10b1019S}
\bibinfo{author}{\bibfnamefont{L.}~\bibnamefont{{S{\'a}}}},
  \bibinfo{author}{\bibfnamefont{P.}~\bibnamefont{{Ribeiro}}},
  \bibnamefont{and} \bibinfo{author}{\bibfnamefont{T.}~\bibnamefont{{Prosen}}},
  \bibinfo{journal}{Physical Review X} \textbf{\bibinfo{volume}{10}},
  \bibinfo{eid}{021019} (\bibinfo{year}{2020}{\natexlab{a}}),
  \eprint{1910.12784}.

\bibitem[{\citenamefont{{Li} et~al.}(2021)\citenamefont{{Li}, {Prosen}, and
  {Chan}}}]{2021PhRvL.127q0602L}
\bibinfo{author}{\bibfnamefont{J.}~\bibnamefont{{Li}}},
  \bibinfo{author}{\bibfnamefont{T.}~\bibnamefont{{Prosen}}}, \bibnamefont{and}
  \bibinfo{author}{\bibfnamefont{A.}~\bibnamefont{{Chan}}},
  \bibinfo{journal}{\prl} \textbf{\bibinfo{volume}{127}}, \bibinfo{eid}{170602}
  (\bibinfo{year}{2021}), \eprint{2103.05001}.

\bibitem[{\citenamefont{{S{\'a}} et~al.}(2021)\citenamefont{{S{\'a}},
  {Ribeiro}, and {Prosen}}}]{2021arXiv211212109S}
\bibinfo{author}{\bibfnamefont{L.}~\bibnamefont{{S{\'a}}}},
  \bibinfo{author}{\bibfnamefont{P.}~\bibnamefont{{Ribeiro}}},
  \bibnamefont{and} \bibinfo{author}{\bibfnamefont{T.}~\bibnamefont{{Prosen}}},
  \bibinfo{journal}{arXiv e-prints} \bibinfo{eid}{arXiv:2112.12109}
  (\bibinfo{year}{2021}), \eprint{2112.12109}.

\bibitem[{\citenamefont{{Kulkarni} et~al.}(2021)\citenamefont{{Kulkarni},
  {Numasawa}, and {Ryu}}}]{2021arXiv211213489K}
\bibinfo{author}{\bibfnamefont{A.}~\bibnamefont{{Kulkarni}}},
  \bibinfo{author}{\bibfnamefont{T.}~\bibnamefont{{Numasawa}}},
  \bibnamefont{and} \bibinfo{author}{\bibfnamefont{S.}~\bibnamefont{{Ryu}}},
  \bibinfo{journal}{arXiv e-prints} \bibinfo{eid}{arXiv:2112.13489}
  (\bibinfo{year}{2021}), \eprint{2112.13489}.

\bibitem[{\citenamefont{{Denisov} et~al.}(2019)\citenamefont{{Denisov},
  {Laptyeva}, {Tarnowski}, {Chru{\'s}ci{\'n}ski}, and
  {{\.Z}yczkowski}}}]{2019PhRvL.123n0403D}
\bibinfo{author}{\bibfnamefont{S.}~\bibnamefont{{Denisov}}},
  \bibinfo{author}{\bibfnamefont{T.}~\bibnamefont{{Laptyeva}}},
  \bibinfo{author}{\bibfnamefont{W.}~\bibnamefont{{Tarnowski}}},
  \bibinfo{author}{\bibfnamefont{D.}~\bibnamefont{{Chru{\'s}ci{\'n}ski}}},
  \bibnamefont{and}
  \bibinfo{author}{\bibfnamefont{K.}~\bibnamefont{{{\.Z}yczkowski}}},
  \bibinfo{journal}{\prl} \textbf{\bibinfo{volume}{123}}, \bibinfo{eid}{140403}
  (\bibinfo{year}{2019}), \eprint{1811.12282}.

\bibitem[{\citenamefont{{Wang} et~al.}(2020)\citenamefont{{Wang}, {Piazza}, and
  {Luitz}}}]{2020PhRvL.124j0604W}
\bibinfo{author}{\bibfnamefont{K.}~\bibnamefont{{Wang}}},
  \bibinfo{author}{\bibfnamefont{F.}~\bibnamefont{{Piazza}}}, \bibnamefont{and}
  \bibinfo{author}{\bibfnamefont{D.~J.} \bibnamefont{{Luitz}}},
  \bibinfo{journal}{\prl} \textbf{\bibinfo{volume}{124}}, \bibinfo{eid}{100604}
  (\bibinfo{year}{2020}), \eprint{1911.05740}.

\bibitem[{\citenamefont{{Sommer} et~al.}(2021)\citenamefont{{Sommer}, {Piazza},
  and {Luitz}}}]{2021PhRvR...3b3190S}
\bibinfo{author}{\bibfnamefont{O.~E.} \bibnamefont{{Sommer}}},
  \bibinfo{author}{\bibfnamefont{F.}~\bibnamefont{{Piazza}}}, \bibnamefont{and}
  \bibinfo{author}{\bibfnamefont{D.~J.} \bibnamefont{{Luitz}}},
  \bibinfo{journal}{Physical Review Research} \textbf{\bibinfo{volume}{3}},
  \bibinfo{eid}{023190} (\bibinfo{year}{2021}), \eprint{2011.08853}.

\bibitem[{\citenamefont{{Ginibre}}(1965)}]{1965JMP.....6..440G}
\bibinfo{author}{\bibfnamefont{J.}~\bibnamefont{{Ginibre}}},
  \bibinfo{journal}{Journal of Mathematical Physics}
  \textbf{\bibinfo{volume}{6}}, \bibinfo{pages}{440} (\bibinfo{year}{1965}).

\bibitem[{\citenamefont{{Page}}(1993{\natexlab{a}})}]{1993PhRvL..71.1291P}
\bibinfo{author}{\bibfnamefont{D.~N.} \bibnamefont{{Page}}},
  \bibinfo{journal}{\prl} \textbf{\bibinfo{volume}{71}}, \bibinfo{pages}{1291}
  (\bibinfo{year}{1993}{\natexlab{a}}), \eprint{gr-qc/9305007}.

\bibitem[{\citenamefont{{Garc{\'\i}a-Garc{\'\i}a}
  et~al.}(2022)\citenamefont{{Garc{\'\i}a-Garc{\'\i}a}, {S{\'a}}, and
  {Verbaarschot}}}]{2022PhRvX..12b1040G}
\bibinfo{author}{\bibfnamefont{A.~M.} \bibnamefont{{Garc{\'\i}a-Garc{\'\i}a}}},
  \bibinfo{author}{\bibfnamefont{L.}~\bibnamefont{{S{\'a}}}}, \bibnamefont{and}
  \bibinfo{author}{\bibfnamefont{J.~J.~M.} \bibnamefont{{Verbaarschot}}},
  \bibinfo{journal}{Physical Review X} \textbf{\bibinfo{volume}{12}},
  \bibinfo{eid}{021040} (\bibinfo{year}{2022}), \eprint{2110.03444}.

\bibitem[{\citenamefont{{Shivam} et~al.}(2022)\citenamefont{{Shivam}, {De
  Luca}, {Huse}, and {Chan}}}]{Chan_forthcoming}
\bibinfo{author}{\bibfnamefont{S.}~\bibnamefont{{Shivam}}},
  \bibinfo{author}{\bibfnamefont{A.}~\bibnamefont{{De Luca}}},
  \bibinfo{author}{\bibfnamefont{D.~A.} \bibnamefont{{Huse}}},
  \bibnamefont{and} \bibinfo{author}{\bibfnamefont{A.}~\bibnamefont{{Chan}}},
  \bibinfo{journal}{arXiv e-prints} \bibinfo{eid}{arXiv:2207.12390}
  (\bibinfo{year}{2022}), \eprint{2207.12390}.

\bibitem[{\citenamefont{Grobe et~al.}(1988)\citenamefont{Grobe, Haake, and
  Sommers}}]{PhysRevLett.61.1899}
\bibinfo{author}{\bibfnamefont{R.}~\bibnamefont{Grobe}},
  \bibinfo{author}{\bibfnamefont{F.}~\bibnamefont{Haake}}, \bibnamefont{and}
  \bibinfo{author}{\bibfnamefont{H.-J.} \bibnamefont{Sommers}},
  \bibinfo{journal}{Phys. Rev. Lett.} \textbf{\bibinfo{volume}{61}},
  \bibinfo{pages}{1899} (\bibinfo{year}{1988}),
  \urlprefix\url{https://link.aps.org/doi/10.1103/PhysRevLett.61.1899}.

\bibitem[{\citenamefont{Grobe and Haake}(1989)}]{PhysRevLett.62.2893}
\bibinfo{author}{\bibfnamefont{R.}~\bibnamefont{Grobe}} \bibnamefont{and}
  \bibinfo{author}{\bibfnamefont{F.}~\bibnamefont{Haake}},
  \bibinfo{journal}{Phys. Rev. Lett.} \textbf{\bibinfo{volume}{62}},
  \bibinfo{pages}{2893} (\bibinfo{year}{1989}),
  \urlprefix\url{https://link.aps.org/doi/10.1103/PhysRevLett.62.2893}.

\bibitem[{\citenamefont{{Brody}}(2014)}]{2014JPhA...47c5305B}
\bibinfo{author}{\bibfnamefont{D.~C.} \bibnamefont{{Brody}}},
  \bibinfo{journal}{Journal of Physics A Mathematical General}
  \textbf{\bibinfo{volume}{47}}, \bibinfo{eid}{035305} (\bibinfo{year}{2014}),
  \eprint{1308.2609}.

\bibitem[{Note1()}]{Note1}
Note1, \bibinfo{note}{$-\protect \text {Tr}\rho \protect \qopname \relax
  o{log}\rho $ becomes ambiguous due to the complex arguments of the logarithm.
  Choosing the principal value, the answer becomes $O(N_A)$ which we do not
  expect to be useful (see Supplemental Material, which includes \cite
  {1962JMP.....3.1191D, 2009arXiv0907.5605E,
  bourgade2018random,2013arXiv1312.1301B, 2021arXiv210306730C,
  2020arXiv200508425M, 2022arXiv22XXXXXXXC2}).}

\bibitem[{\citenamefont{{Page}}(1993{\natexlab{b}})}]{1993PhRvL..71.3743P}
\bibinfo{author}{\bibfnamefont{D.~N.} \bibnamefont{{Page}}},
  \bibinfo{journal}{\prl} \textbf{\bibinfo{volume}{71}}, \bibinfo{pages}{3743}
  (\bibinfo{year}{1993}{\natexlab{b}}), \eprint{hep-th/9306083}.

\bibitem[{\citenamefont{{Vidmar} and {Rigol}}(2017)}]{2017PhRvL.119v0603V}
\bibinfo{author}{\bibfnamefont{L.}~\bibnamefont{{Vidmar}}} \bibnamefont{and}
  \bibinfo{author}{\bibfnamefont{M.}~\bibnamefont{{Rigol}}},
  \bibinfo{journal}{\prl} \textbf{\bibinfo{volume}{119}}, \bibinfo{eid}{220603}
  (\bibinfo{year}{2017}), \eprint{1708.08453}.

\bibitem[{\citenamefont{{Penington}}(2020)}]{2020JHEP...09..002P}
\bibinfo{author}{\bibfnamefont{G.}~\bibnamefont{{Penington}}},
  \bibinfo{journal}{Journal of High Energy Physics}
  \textbf{\bibinfo{volume}{2020}}, \bibinfo{eid}{2} (\bibinfo{year}{2020}),
  \eprint{1905.08255}.

\bibitem[{\citenamefont{{Almheiri} et~al.}(2019)\citenamefont{{Almheiri},
  {Engelhardt}, {Marolf}, and {Maxfield}}}]{2019JHEP...12..063A}
\bibinfo{author}{\bibfnamefont{A.}~\bibnamefont{{Almheiri}}},
  \bibinfo{author}{\bibfnamefont{N.}~\bibnamefont{{Engelhardt}}},
  \bibinfo{author}{\bibfnamefont{D.}~\bibnamefont{{Marolf}}}, \bibnamefont{and}
  \bibinfo{author}{\bibfnamefont{H.}~\bibnamefont{{Maxfield}}},
  \bibinfo{journal}{Journal of High Energy Physics}
  \textbf{\bibinfo{volume}{2019}}, \bibinfo{eid}{63} (\bibinfo{year}{2019}),
  \eprint{1905.08762}.

\bibitem[{\citenamefont{Wishart}(1928)}]{wishart1928generalised}
\bibinfo{author}{\bibfnamefont{J.}~\bibnamefont{Wishart}},
  \bibinfo{journal}{Biometrika}  (\bibinfo{year}{1928}).

\bibitem[{\citenamefont{{Feinberg} and {Zee}}(1997)}]{1997NuPhB.504..579F}
\bibinfo{author}{\bibfnamefont{J.}~\bibnamefont{{Feinberg}}} \bibnamefont{and}
  \bibinfo{author}{\bibfnamefont{A.}~\bibnamefont{{Zee}}},
  \bibinfo{journal}{Nuclear Physics B} \textbf{\bibinfo{volume}{504}},
  \bibinfo{pages}{579} (\bibinfo{year}{1997}), \eprint{cond-mat/9703087}.

\bibitem[{\citenamefont{Bourgade and Yau}(2017)}]{2013arXiv1312.1301B}
\bibinfo{author}{\bibfnamefont{P.}~\bibnamefont{Bourgade}} \bibnamefont{and}
  \bibinfo{author}{\bibfnamefont{H.-T.} \bibnamefont{Yau}},
  \bibinfo{journal}{Communications in Mathematical Physics}
  \textbf{\bibinfo{volume}{350}}, \bibinfo{pages}{231} (\bibinfo{year}{2017}).

\bibitem[{\citenamefont{{Marcinek} and {Yau}}(2020)}]{2020arXiv200508425M}
\bibinfo{author}{\bibfnamefont{J.}~\bibnamefont{{Marcinek}}} \bibnamefont{and}
  \bibinfo{author}{\bibfnamefont{H.-T.} \bibnamefont{{Yau}}},
  \bibinfo{journal}{arXiv e-prints} \bibinfo{eid}{arXiv:2005.08425}
  (\bibinfo{year}{2020}), \eprint{2005.08425}.

\bibitem[{\citenamefont{{Benigni} and {Lopatto}}(2022)}]{2022CMaPh.391..401B}
\bibinfo{author}{\bibfnamefont{L.}~\bibnamefont{{Benigni}}} \bibnamefont{and}
  \bibinfo{author}{\bibfnamefont{P.}~\bibnamefont{{Lopatto}}},
  \bibinfo{journal}{Communications in Mathematical Physics}
  \textbf{\bibinfo{volume}{391}}, \bibinfo{pages}{401} (\bibinfo{year}{2022}),
  \eprint{2103.12013}.

\bibitem[{\citenamefont{Cipolloni et~al.}(2022)\citenamefont{Cipolloni, Erdős,
  and Schröder}}]{2021arXiv210306730C}
\bibinfo{author}{\bibfnamefont{G.}~\bibnamefont{Cipolloni}},
  \bibinfo{author}{\bibfnamefont{L.}~\bibnamefont{Erdős}}, \bibnamefont{and}
  \bibinfo{author}{\bibfnamefont{D.}~\bibnamefont{Schröder}},
  \bibinfo{journal}{The Annals of Probability} \textbf{\bibinfo{volume}{50}},
  \bibinfo{pages}{984 } (\bibinfo{year}{2022}),
  \urlprefix\url{https://doi.org/10.1214/21-AOP1552}.

\bibitem[{\citenamefont{{Cipolloni} et~al.}(2022)\citenamefont{{Cipolloni},
  {Erd{\H{o}}s}, and {Schr{\"o}der}}}]{2022arXiv220301861C}
\bibinfo{author}{\bibfnamefont{G.}~\bibnamefont{{Cipolloni}}},
  \bibinfo{author}{\bibfnamefont{L.}~\bibnamefont{{Erd{\H{o}}s}}},
  \bibnamefont{and}
  \bibinfo{author}{\bibfnamefont{D.}~\bibnamefont{{Schr{\"o}der}}},
  \bibinfo{journal}{arXiv e-prints} \bibinfo{eid}{arXiv:2203.01861}
  (\bibinfo{year}{2022}), \eprint{2203.01861}.

\bibitem[{\citenamefont{Bourgade et~al.}(2018)\citenamefont{Bourgade, Yau, and
  Yin}}]{bourgade2018random}
\bibinfo{author}{\bibfnamefont{P.}~\bibnamefont{Bourgade}},
  \bibinfo{author}{\bibfnamefont{H.-T.} \bibnamefont{Yau}}, \bibnamefont{and}
  \bibinfo{author}{\bibfnamefont{J.}~\bibnamefont{Yin}},
  \bibinfo{journal}{arXiv preprint arXiv:1807.01559}  (\bibinfo{year}{2018}).

\bibitem[{\citenamefont{Collins}(2003)}]{2002math.ph...5010C}
\bibinfo{author}{\bibfnamefont{B.}~\bibnamefont{Collins}},
  \bibinfo{journal}{International Mathematics Research Notices}
  \textbf{\bibinfo{volume}{2003}}, \bibinfo{pages}{953} (\bibinfo{year}{2003}).

\bibitem[{\citenamefont{{Fyodorov}}(2018)}]{2018CMaPh.363..579F}
\bibinfo{author}{\bibfnamefont{Y.~V.} \bibnamefont{{Fyodorov}}},
  \bibinfo{journal}{Communications in Mathematical Physics}
  \textbf{\bibinfo{volume}{363}}, \bibinfo{pages}{579} (\bibinfo{year}{2018}),
  \eprint{1710.04699}.

\bibitem[{Note2()}]{Note2}
Note2, \bibinfo{note}{note that if we had taken the density matrix to be
  Hermitian i.e.~$\mathinner {|{R}\rangle }\mathinner {\langle {R}|}$ or
  $\mathinner {|{L}\rangle }\mathinner {\langle {L}|}$, this would imply that
  the reduced density matrices are, once again, drawn from the Wishart
  ensemble.}

\bibitem[{\citenamefont{Bourgade and Dubach}(2020)}]{2018arXiv180101219B}
\bibinfo{author}{\bibfnamefont{P.}~\bibnamefont{Bourgade}} \bibnamefont{and}
  \bibinfo{author}{\bibfnamefont{G.}~\bibnamefont{Dubach}},
  \bibinfo{journal}{Probability Theory and Related Fields}
  \textbf{\bibinfo{volume}{177}}, \bibinfo{pages}{397} (\bibinfo{year}{2020}).

\bibitem[{\citenamefont{{Akemann}}(2011)}]{2011arXiv1104.5203A}
\bibinfo{author}{\bibfnamefont{G.}~\bibnamefont{{Akemann}}},
  \bibinfo{journal}{arXiv e-prints} \bibinfo{eid}{arXiv:1104.5203}
  (\bibinfo{year}{2011}), \eprint{1104.5203}.

\bibitem[{\citenamefont{{Burda} et~al.}(2010)\citenamefont{{Burda}, {Jarosz},
  {Livan}, {Nowak}, and {Swiech}}}]{2010PhRvE..82f1114B}
\bibinfo{author}{\bibfnamefont{Z.}~\bibnamefont{{Burda}}},
  \bibinfo{author}{\bibfnamefont{A.}~\bibnamefont{{Jarosz}}},
  \bibinfo{author}{\bibfnamefont{G.}~\bibnamefont{{Livan}}},
  \bibinfo{author}{\bibfnamefont{M.~A.} \bibnamefont{{Nowak}}},
  \bibnamefont{and} \bibinfo{author}{\bibfnamefont{A.}~\bibnamefont{{Swiech}}},
  \bibinfo{journal}{\pre} \textbf{\bibinfo{volume}{82}}, \bibinfo{eid}{061114}
  (\bibinfo{year}{2010}), \eprint{1007.3594}.

\bibitem[{\citenamefont{{Akemann} et~al.}(2021)\citenamefont{{Akemann}, {Byun},
  and {Kang}}}]{2021AnHP...22.1035A}
\bibinfo{author}{\bibfnamefont{G.}~\bibnamefont{{Akemann}}},
  \bibinfo{author}{\bibfnamefont{S.-S.} \bibnamefont{{Byun}}},
  \bibnamefont{and} \bibinfo{author}{\bibfnamefont{N.-G.}
  \bibnamefont{{Kang}}}, \bibinfo{journal}{Annales Henri Poincare;}
  \textbf{\bibinfo{volume}{22}}, \bibinfo{pages}{1035} (\bibinfo{year}{2021}),
  \eprint{2004.07626}.

\bibitem[{\citenamefont{Rider and Vir{\'a}g}(2007)}]{2006math......6663R}
\bibinfo{author}{\bibfnamefont{B.}~\bibnamefont{Rider}} \bibnamefont{and}
  \bibinfo{author}{\bibfnamefont{B.}~\bibnamefont{Vir{\'a}g}},
  \bibinfo{journal}{International Mathematics Research Notices}
  \textbf{\bibinfo{volume}{2007}} (\bibinfo{year}{2007}).

\bibitem[{\citenamefont{{Hamazaki} et~al.}(2020)\citenamefont{{Hamazaki},
  {Kawabata}, {Kura}, and {Ueda}}}]{2020PhRvR...2b3286H}
\bibinfo{author}{\bibfnamefont{R.}~\bibnamefont{{Hamazaki}}},
  \bibinfo{author}{\bibfnamefont{K.}~\bibnamefont{{Kawabata}}},
  \bibinfo{author}{\bibfnamefont{N.}~\bibnamefont{{Kura}}}, \bibnamefont{and}
  \bibinfo{author}{\bibfnamefont{M.}~\bibnamefont{{Ueda}}},
  \bibinfo{journal}{Physical Review Research} \textbf{\bibinfo{volume}{2}},
  \bibinfo{eid}{023286} (\bibinfo{year}{2020}), \eprint{1904.13082}.

\bibitem[{\citenamefont{Deutsch}(1991)}]{PhysRevA.43.2046}
\bibinfo{author}{\bibfnamefont{J.~M.} \bibnamefont{Deutsch}},
  \bibinfo{journal}{Phys. Rev. A} \textbf{\bibinfo{volume}{43}},
  \bibinfo{pages}{2046} (\bibinfo{year}{1991}),
  \urlprefix\url{https://link.aps.org/doi/10.1103/PhysRevA.43.2046}.

\bibitem[{\citenamefont{{Srednicki}}(1994)}]{1994PhRvE..50..888S}
\bibinfo{author}{\bibfnamefont{M.}~\bibnamefont{{Srednicki}}},
  \bibinfo{journal}{\pre} \textbf{\bibinfo{volume}{50}}, \bibinfo{pages}{888}
  (\bibinfo{year}{1994}), \eprint{cond-mat/9403051}.

\bibitem[{\citenamefont{Cipolloni et~al.}(2021)\citenamefont{Cipolloni,
  Erd{\H{o}}s, and Schr{\"o}der}}]{cipolloni2021eigenstate}
\bibinfo{author}{\bibfnamefont{G.}~\bibnamefont{Cipolloni}},
  \bibinfo{author}{\bibfnamefont{L.}~\bibnamefont{Erd{\H{o}}s}},
  \bibnamefont{and}
  \bibinfo{author}{\bibfnamefont{D.}~\bibnamefont{Schr{\"o}der}},
  \bibinfo{journal}{Communications in Mathematical Physics}
  \textbf{\bibinfo{volume}{388}}, \bibinfo{pages}{1005} (\bibinfo{year}{2021}).

\bibitem[{\citenamefont{{Cipolloni} and
  {Kudler-Flam}}(2022)}]{2022arXiv22XXXXXXXC}
\bibinfo{author}{\bibfnamefont{G.}~\bibnamefont{{Cipolloni}}} \bibnamefont{and}
  \bibinfo{author}{\bibfnamefont{J.}~\bibnamefont{{Kudler-Flam}}},
  \bibinfo{journal}{arXiv e-prints} \bibinfo{eid}{arXiv:22XX.xxxxx}
  (\bibinfo{year}{2022}), \eprint{22XX.xxxxx}.

\bibitem[{\citenamefont{{Modak} and {Mandal}}(2021)}]{2021PhRvA.103f2416M}
\bibinfo{author}{\bibfnamefont{R.}~\bibnamefont{{Modak}}} \bibnamefont{and}
  \bibinfo{author}{\bibfnamefont{B.~P.} \bibnamefont{{Mandal}}},
  \bibinfo{journal}{\pra} \textbf{\bibinfo{volume}{103}}, \bibinfo{eid}{062416}
  (\bibinfo{year}{2021}), \eprint{2102.01097}.

\bibitem[{\citenamefont{{Can}}(2019)}]{2019JPhA...52V5302C}
\bibinfo{author}{\bibfnamefont{T.}~\bibnamefont{{Can}}},
  \bibinfo{journal}{Journal of Physics A Mathematical General}
  \textbf{\bibinfo{volume}{52}}, \bibinfo{eid}{485302} (\bibinfo{year}{2019}),
  \eprint{1902.01442}.

\bibitem[{\citenamefont{{Can} et~al.}(2019)\citenamefont{{Can}, {Oganesyan},
  {Orgad}, and {Gopalakrishnan}}}]{2019PhRvL.123w4103C}
\bibinfo{author}{\bibfnamefont{T.}~\bibnamefont{{Can}}},
  \bibinfo{author}{\bibfnamefont{V.}~\bibnamefont{{Oganesyan}}},
  \bibinfo{author}{\bibfnamefont{D.}~\bibnamefont{{Orgad}}}, \bibnamefont{and}
  \bibinfo{author}{\bibfnamefont{S.}~\bibnamefont{{Gopalakrishnan}}},
  \bibinfo{journal}{\prl} \textbf{\bibinfo{volume}{123}}, \bibinfo{eid}{234103}
  (\bibinfo{year}{2019}), \eprint{1902.01414}.

\bibitem[{\citenamefont{{S{\'a}}
  et~al.}(2020{\natexlab{b}})\citenamefont{{S{\'a}}, {Ribeiro}, and
  {Prosen}}}]{2020JPhA...53D5303S}
\bibinfo{author}{\bibfnamefont{L.}~\bibnamefont{{S{\'a}}}},
  \bibinfo{author}{\bibfnamefont{P.}~\bibnamefont{{Ribeiro}}},
  \bibnamefont{and} \bibinfo{author}{\bibfnamefont{T.}~\bibnamefont{{Prosen}}},
  \bibinfo{journal}{Journal of Physics A Mathematical General}
  \textbf{\bibinfo{volume}{53}}, \bibinfo{eid}{305303}
  (\bibinfo{year}{2020}{\natexlab{b}}), \eprint{1905.02155}.

\bibitem[{\citenamefont{{Lange} and {Timm}}(2021)}]{2021Chaos..31b3101L}
\bibinfo{author}{\bibfnamefont{S.}~\bibnamefont{{Lange}}} \bibnamefont{and}
  \bibinfo{author}{\bibfnamefont{C.}~\bibnamefont{{Timm}}},
  \bibinfo{journal}{Chaos} \textbf{\bibinfo{volume}{31}}, \bibinfo{eid}{023101}
  (\bibinfo{year}{2021}).

\bibitem[{\citenamefont{{Tarnowski} et~al.}(2021)\citenamefont{{Tarnowski},
  {Yusipov}, {Laptyeva}, {Denisov}, {Chru{\'s}ci{\'n}ski}, and
  {{\.Z}yczkowski}}}]{2021PhRvE.104c4118T}
\bibinfo{author}{\bibfnamefont{W.}~\bibnamefont{{Tarnowski}}},
  \bibinfo{author}{\bibfnamefont{I.}~\bibnamefont{{Yusipov}}},
  \bibinfo{author}{\bibfnamefont{T.}~\bibnamefont{{Laptyeva}}},
  \bibinfo{author}{\bibfnamefont{S.}~\bibnamefont{{Denisov}}},
  \bibinfo{author}{\bibfnamefont{D.}~\bibnamefont{{Chru{\'s}ci{\'n}ski}}},
  \bibnamefont{and}
  \bibinfo{author}{\bibfnamefont{K.}~\bibnamefont{{{\.Z}yczkowski}}},
  \bibinfo{journal}{\pre} \textbf{\bibinfo{volume}{104}}, \bibinfo{eid}{034118}
  (\bibinfo{year}{2021}), \eprint{2105.02369}.

\bibitem[{\citenamefont{{Turkeshi} and
  {Schir{\'o}}}(2022)}]{2022arXiv220109895T}
\bibinfo{author}{\bibfnamefont{X.}~\bibnamefont{{Turkeshi}}} \bibnamefont{and}
  \bibinfo{author}{\bibfnamefont{M.}~\bibnamefont{{Schir{\'o}}}},
  \bibinfo{journal}{arXiv e-prints} \bibinfo{eid}{arXiv:2201.09895}
  (\bibinfo{year}{2022}), \eprint{2201.09895}.

\bibitem[{\citenamefont{{Kawabata} et~al.}(2022)\citenamefont{{Kawabata},
  {Numasawa}, and {Ryu}}}]{2022arXiv220605384K}
\bibinfo{author}{\bibfnamefont{K.}~\bibnamefont{{Kawabata}}},
  \bibinfo{author}{\bibfnamefont{T.}~\bibnamefont{{Numasawa}}},
  \bibnamefont{and} \bibinfo{author}{\bibfnamefont{S.}~\bibnamefont{{Ryu}}},
  \bibinfo{journal}{arXiv e-prints} \bibinfo{eid}{arXiv:2206.05384}
  (\bibinfo{year}{2022}), \eprint{2206.05384}.

\bibitem[{\citenamefont{{Pakrouski} et~al.}(2021)\citenamefont{{Pakrouski},
  {Pallegar}, {Popov}, and {Klebanov}}}]{2021PhRvR...3d3156P}
\bibinfo{author}{\bibfnamefont{K.}~\bibnamefont{{Pakrouski}}},
  \bibinfo{author}{\bibfnamefont{P.~N.} \bibnamefont{{Pallegar}}},
  \bibinfo{author}{\bibfnamefont{F.~K.} \bibnamefont{{Popov}}},
  \bibnamefont{and} \bibinfo{author}{\bibfnamefont{I.~R.}
  \bibnamefont{{Klebanov}}}, \bibinfo{journal}{Physical Review Research}
  \textbf{\bibinfo{volume}{3}}, \bibinfo{eid}{043156} (\bibinfo{year}{2021}),
  \eprint{2106.10300}.

\bibitem[{\citenamefont{{Weinberg} and {Bukov}}(2017)}]{2017ScPP....2....3W}
\bibinfo{author}{\bibfnamefont{P.}~\bibnamefont{{Weinberg}}} \bibnamefont{and}
  \bibinfo{author}{\bibfnamefont{M.}~\bibnamefont{{Bukov}}},
  \bibinfo{journal}{SciPost Physics} \textbf{\bibinfo{volume}{2}},
  \bibinfo{eid}{003} (\bibinfo{year}{2017}), \eprint{1610.03042}.

\bibitem[{\citenamefont{{Weinberg} and {Bukov}}(2019)}]{2019ScPP....7...20W}
\bibinfo{author}{\bibfnamefont{P.}~\bibnamefont{{Weinberg}}} \bibnamefont{and}
  \bibinfo{author}{\bibfnamefont{M.}~\bibnamefont{{Bukov}}},
  \bibinfo{journal}{SciPost Physics} \textbf{\bibinfo{volume}{7}},
  \bibinfo{eid}{020} (\bibinfo{year}{2019}), \eprint{1804.06782}.

\bibitem[{\citenamefont{{Dyson}}(1962)}]{1962JMP.....3.1191D}
\bibinfo{author}{\bibfnamefont{F.~J.} \bibnamefont{{Dyson}}},
  \bibinfo{journal}{Journal of Mathematical Physics}
  \textbf{\bibinfo{volume}{3}}, \bibinfo{pages}{1191} (\bibinfo{year}{1962}).

\bibitem[{\citenamefont{{Erdos} et~al.}(2009)\citenamefont{{Erdos}, {Schlein},
  and {Yau}}}]{2009arXiv0907.5605E}
\bibinfo{author}{\bibfnamefont{L.}~\bibnamefont{{Erdos}}},
  \bibinfo{author}{\bibfnamefont{B.}~\bibnamefont{{Schlein}}},
  \bibnamefont{and} \bibinfo{author}{\bibfnamefont{H.-T.} \bibnamefont{{Yau}}},
  \bibinfo{journal}{arXiv e-prints} \bibinfo{eid}{arXiv:0907.5605}
  (\bibinfo{year}{2009}), \eprint{0907.5605}.

\bibitem[{\citenamefont{{Benigni} and
  {Cipolloni}}(2022)}]{2022arXiv22XXXXXXXC2}
\bibinfo{author}{\bibfnamefont{L.}~\bibnamefont{{Benigni}}} \bibnamefont{and}
  \bibinfo{author}{\bibfnamefont{G.}~\bibnamefont{{Cipolloni}}},
  \bibinfo{journal}{arXiv e-prints} \bibinfo{eid}{arXiv:22XX.xxxxx}
  (\bibinfo{year}{2022}), \eprint{22XX.xxxxx}.

\end{thebibliography}

\pagebreak
\widetext
\begin{center}
\textbf{\large Supplemental Materials}
\end{center}

\section{Dyson Brownian Motion}
Consider an $N\times N$ Hermitian matrix $H$ evolving under diffusion, i.e. $H(t)=H+B(t)$, with $B(t)$ being a Hermitian matrix valued Brownian motion. Dyson's revolutionary discovery \cite{1962JMP.....3.1191D} was that the eigenvalues of $H(t)$ satisfy an autonomous system of stochastic differential equations (SDEs) that do not involve eigenvectors:
\begin{equation}
\label{eq:evaldbm}
\mathrm{d}\lambda_i(t)=\mathrm{d}b_i(t)+\sum_{j\ne i}\frac{1}{\lambda_i(t)-\lambda_j(t)}\, \mathrm{d}t.
\end{equation}
Here $b_i(t)$ is a family of independent identically distributed (i.i.d.) standard real Brownian motions. This system of SDEs is now called the \emph{Dyson Brownian motion (DBM)}. The power of this dynamical approach in studying spectral properties of random matrices has been shown by Erd\H{o}s, Schlein, and Yau in their breakthrough paper \cite{2009arXiv0907.5605E}. In particular, they showed that the dynamic in \eqref{eq:evaldbm} converges to equilibrium very fast (for times $t\sim N^{-1+\epsilon}$). Running the flow \eqref{eq:evaldbm} for this much time is necessary because the dynamics have to run for a time bigger than $N^{-1}$ to forget the initial conditions, since $N^{-1}$ is the typical level spacing of the eigenvalues of $H$.

More recently, these techniques have also been extend to study the eigenvectors of $H(t)$. By second order perturbation theory, it follows that the evolution of the eigenvectors of $H(t)$ is described by (see \cite[Theorem 3.1]{2013arXiv1312.1301B} for the rigorous derivation):
\begin{equation}
\label{eq:evecdbm}
\mathrm{d}{\bf u}_i(t)=\sum_{j\ne i}\frac{\mathrm{d} B_{ij}(t)}{\lambda_i(t)-\lambda_j(t)}{\bf u}_j(t)-\sum_{j\ne i}\frac{1}{(\lambda_i(t)-\lambda_j(t))^2}{\bf u}_i(t)\,\mathrm{d} t.
\end{equation}
Here $B_{ij}(t)$ is a family of i.i.d. standard real Brownian motions independent of the $b_i(t)$'s from \eqref{eq:evaldbm}. The independence of $B_{ij}(t)$ and $b_i(t)$ will play a fundamental role, as explained below \eqref{eq:parpde}.

In this Letter, we are interested in studying the fluctuations of overlaps $\langle {\bf u}_i,A{\bf u}_i\rangle$, with $A\in \mathbb{C}^{N\times N}$ being a deterministic observable matrix. For simplicity of the presentation we now assume that $A$ is Hermitian; at the end of this section we will explain the minor differences required when $A$ is not Hermitian or when also ``off--diagonal'' overlaps $\langle {\bf u}_i,A{\bf u}_j\rangle$ are considered. We remark that the observable matrix $A$ is not required to be of full--rank, i.e. we also allow $\mathrm{rank}(A)\ll N$. Our goal is to prove that
\begin{equation}
\label{eq:gauss}
\sqrt{\frac{N}{\langle AA^*\rangle}}\left[\langle {\bf u}_i,A{\bf u}_i\rangle- \frac{1}{N}\mathrm{Tr} A\right]\sim \mathcal{N}(0,1),
\end{equation}
with $\mathcal{N}(0,1)$ being a standard real Gaussian random variable. To keep the notation simple, from now on and without loss of generality, we assume that $\mathrm{Tr}A=0$ since this can always be achieved by a simple shift in the matrix. For this purpose we compute the moments of $\langle {\bf u}_i,A{\bf u}_i\rangle$ with the goal of showing their convergence (as $N\to \infty$) to Gaussian moments, i.e. for the second moment we compute
\begin{equation}
\label{eq:secmom}
\mathbf{E}\left[\frac{N}{\langle AA^*\rangle}\big|\langle {\bf u}_i,A{\bf u}_i\rangle\big|^2\Bigg|\mathbf{\lambda}\right].
\end{equation}
Here $\mathbf{E}[\cdot|\mathbf{\lambda}]$ denotes the expectation conditioned on the path of the eigenvalues of $H(t)$. In \eqref{eq:secmom} we consider the conditional expectation with respect to the eigenvalues to decouple the randomness of the eigenvectors from the randomness of the eigenvalues; this decoupling is possible since the driving Brownian motions in \eqref{eq:evaldbm}--\eqref{eq:evecdbm} are independent.

The first difficulty in the analysis of \eqref{eq:secmom}, is that the evolution of $\mathbf{E}[|\langle {\bf u}_i,A{\bf u}_i\rangle|^2|\mathbf{\lambda}]$ along the flow \eqref{eq:evecdbm} is not self contained, i.e. it contains off--diagonal overlaps as well. A fundamental observation in \cite{bourgade2018random} was that even though $\mathbf{E}[|\langle {\bf u}_i,A{\bf u}_i\rangle|^2|\mathbf{\lambda}]$ does not have a closed equation, some particular linear combinations (denoted by $f_t$ below) of diagonal and off--diagonal overlaps do. More precisely, define
\begin{equation}
\label{eq:deff}
f_t=f_t(i,j):=\mathbf{E}\left[\frac{N}{(1+2\delta_{ij})\langle AA^*\rangle}\left(\langle {\bf u}_i,A{\bf u}_i\rangle\langle {\bf u}_j,A{\bf u}_j\rangle+2\big|\langle {\bf u}_i,A{\bf u}_j\rangle\big|^2\right)\Bigg|\mathbf{\lambda}\right],
\end{equation}
as a function over indices $i,j\in\mathbb{N}$, then $f_t$ is the solution of the following parabolic partial differential equation (PDE):
\begin{equation}
\label{eq:parpde}
\partial_t f_t(i,j)=\sum_{k\ne i} \frac{f_t(k,j)-f_t(i,j)}{(\lambda_i(t)-\lambda_k(t))^2}+\sum_{k\ne j} \frac{f_t(i,k)-f_t(i,j)}{(\lambda_j(t)-\lambda_k(t))^2}.
\end{equation}
The key fact used in the derivation of \eqref{eq:parpde} is that the driving Brownian motions $B_{ij}(t)$, $b_i(t)$ in \eqref{eq:evaldbm} and \eqref{eq:evecdbm}, respectively, are independent, hence when considering the conditional expectation $\mathbf{E}[\cdot|\mathbf{\lambda}]$ the stochastic term containing $\mathrm{d}B_{ij}(t)$ vanishes.

To analyze the PDE in \eqref{eq:parpde} we use \emph{energy methods}. To keep the presentation clear and not too technical we only explain the main steps in this method and refer the interested reader to \cite{2020arXiv200508425M,2021arXiv210306730C}, where all the technical details are presented. The first fundamental observation is that the main contribution to the dynamics in \eqref{eq:parpde} comes from eigenvectors corresponding to nearby eigenvalues; this is a consequence of the fact that the kernel in \eqref{eq:parpde} is given by $(\lambda_i(t)-\lambda_k(t))^{-2}$, i.e. the kernel is very singular if $i$ and $k$ are close. For this reason we can approximate $f_t\approx g_t$, with $g_t$ being a localized version of $f_t$ over $K$ indices around the fixed indices $i$ and $j$. Here $K$ grows mildly with $N$, i.e. $K=N^{\omega_K}$, for some very small fixed $\omega_K>0$. Then, the energy method consists of two steps:
\begin{enumerate}
    \item Convergence in the $L^2$--sense, i.e. (here $1$ represent the variance of $\mathcal{N}(0,1)$)
    \begin{equation}
    \label{eq:step1}
    \lVert g_t-1\rVert_{L^2} \le K \mathcal{E},
    \end{equation}
    for some small $\mathcal{E}$. However, it still holds $K \mathcal{E}\gg 1$ (for large $N$).
    \item By Nash inequality (see \cite[Propositions 6.24 and 6.29]{2020arXiv200508425M}) it follows (\emph{ultracontractivity})
    \begin{equation}
    \label{eq:step2}
     \lVert g_{2t}-1\rVert_{L^\infty}\le \frac{K}{Nt} \lVert g_t-1\rVert_{L^2}.
    \end{equation}
\end{enumerate}

Finally, combining \eqref{eq:step1}--\eqref{eq:step2}, we conclude
\begin{equation}
\label{eq:step3}
     \lVert g_{2t}-1\rVert_{L^\infty}\le \frac{K}{Nt} \lVert g_t-1\rVert_{L^2}\le \frac{K^2}{Nt} \mathcal{E}\ll 1,
\end{equation}
choosing the time $t$ large enough. More precisely, we can choose $t\sim K^2/N$, which shows that the convergence in \eqref{eq:step3} is very fast, i.e., since $K=N^{\omega_K}$, it is enough to choose a time which is just slightly bigger than $N^{-1}$. Choosing $i=j$ in \eqref{eq:deff}, this shows that
\begin{equation}
\label{eq:rq}
\sqrt{\frac{N}{\langle AA^*\rangle}}\mathbb{E}\big[\langle {\bf u}_i(2t),A{\bf u}_i(2t)\rangle\big]^2\approx1,
\end{equation}
i.e. that the second moment of the eigenvector overlap (after proper normalization) is the same as the one of $\mathcal{N}(0,1)$. Similar computations for higher moments show the convergence in \eqref{eq:gauss}. The fact that this convergence holds at time $t=0$ as well follows by a simple perturbation argument known as \emph{Green's function comparison theorem (GFT)} (see e.g. \cite[Appendix A]{2021arXiv210306730C}).

The analysis to prove Gaussianity (as in \eqref{eq:gauss}) for non--Hermitian $A$'s and for off--diagonal overlaps $\langle {\bf u}_i,A{\bf u}_j\rangle$ is very similar using a slightly more complicated version of the PDE in \eqref{eq:parpde} (see \cite[Theorem 4.8]{2020arXiv200508425M} for the precise equation). As it was shown in \cite{2020arXiv200508425M} for finite ($N$--independent) rank matrices and in \cite{2022arXiv22XXXXXXXC2} for arbitrary $A$'s, the analysis of this equation is technically more complicated but follows exactly the same strategy described in \eqref{eq:step1}--\eqref{eq:rq}.

\section{Additional Data}

\begin{figure}
  \centering
    \includegraphics[width = .48\textwidth]{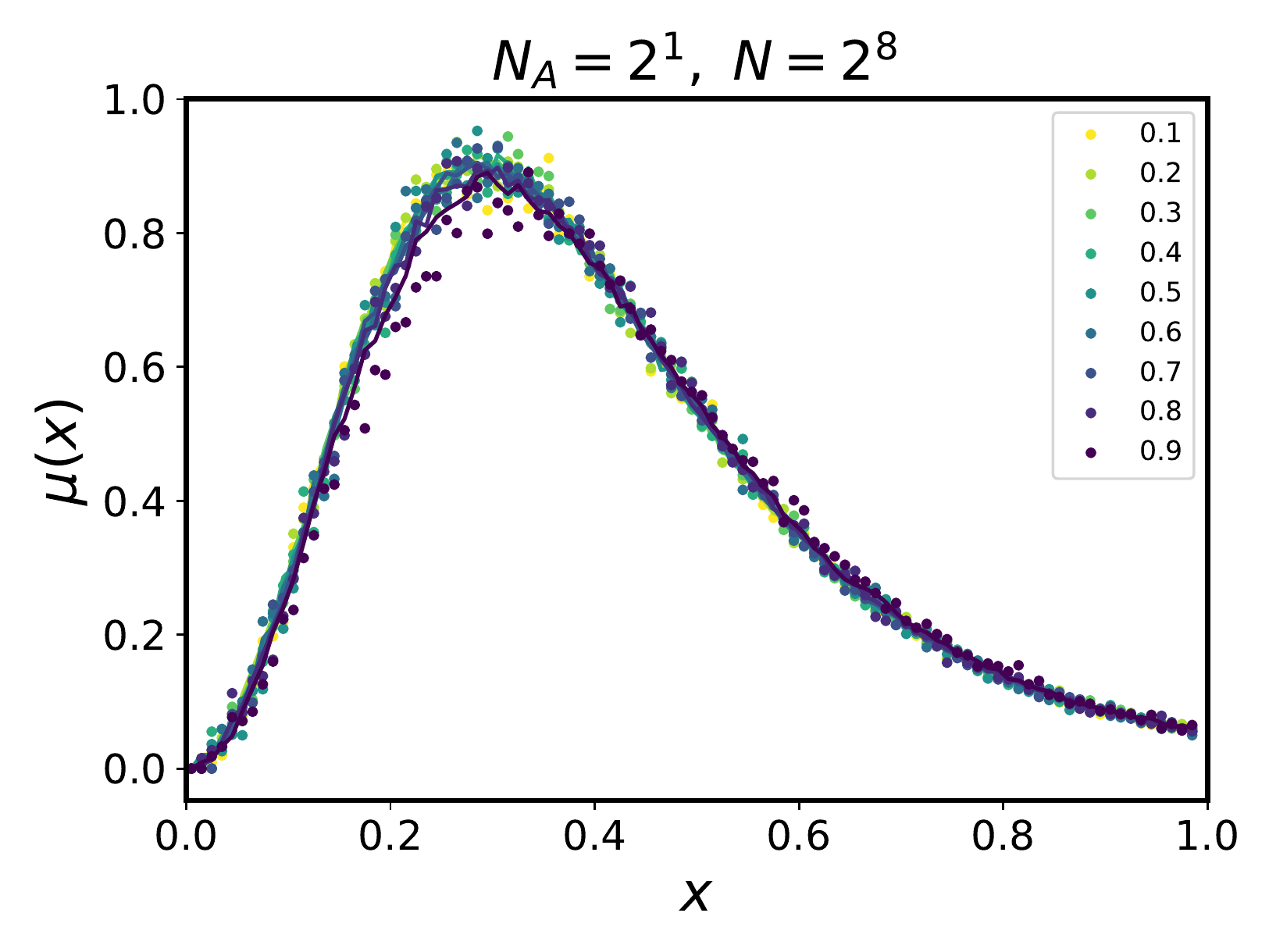}
      \includegraphics[width = .48\textwidth]{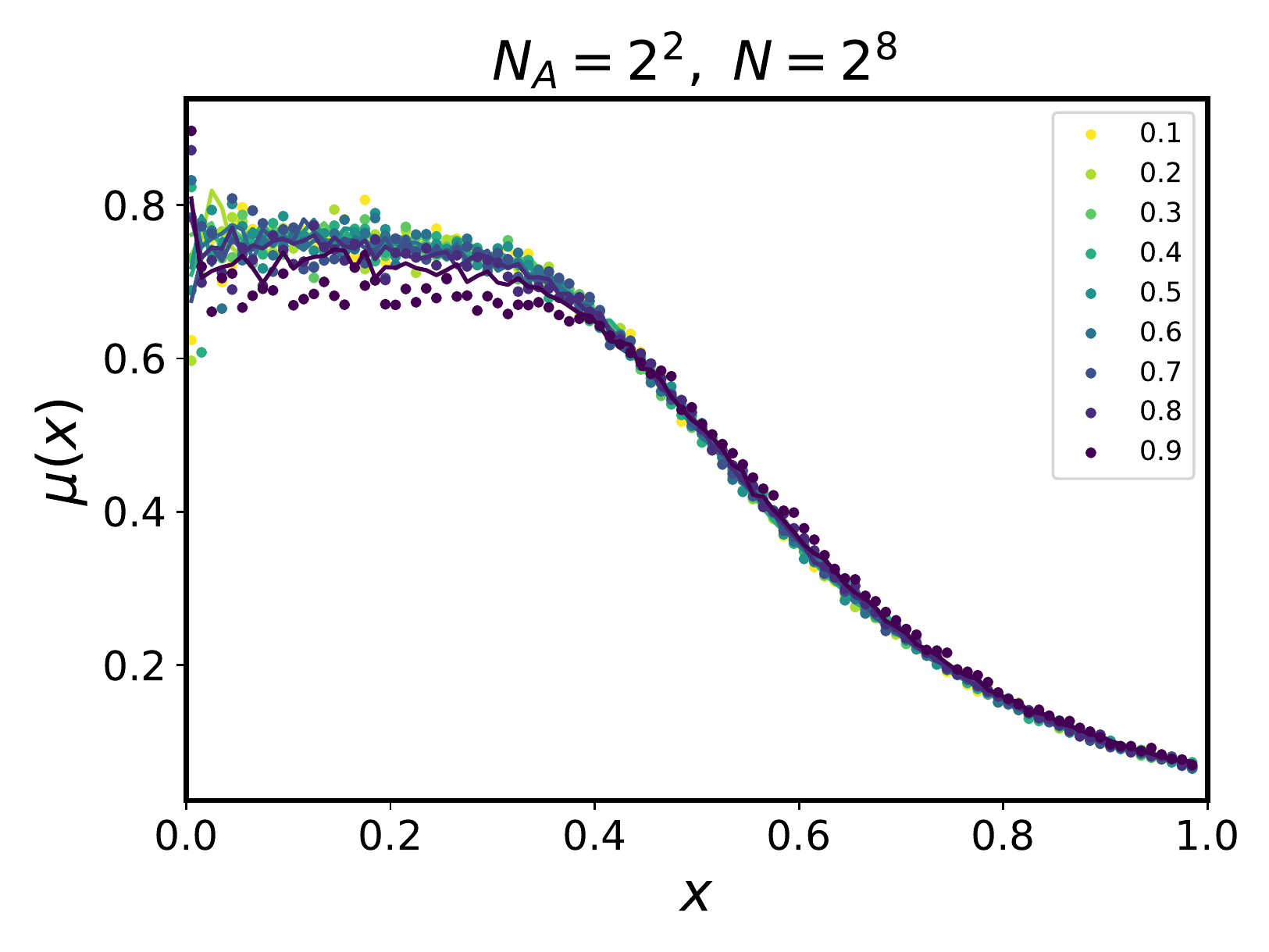}
        \includegraphics[width = .48\textwidth]{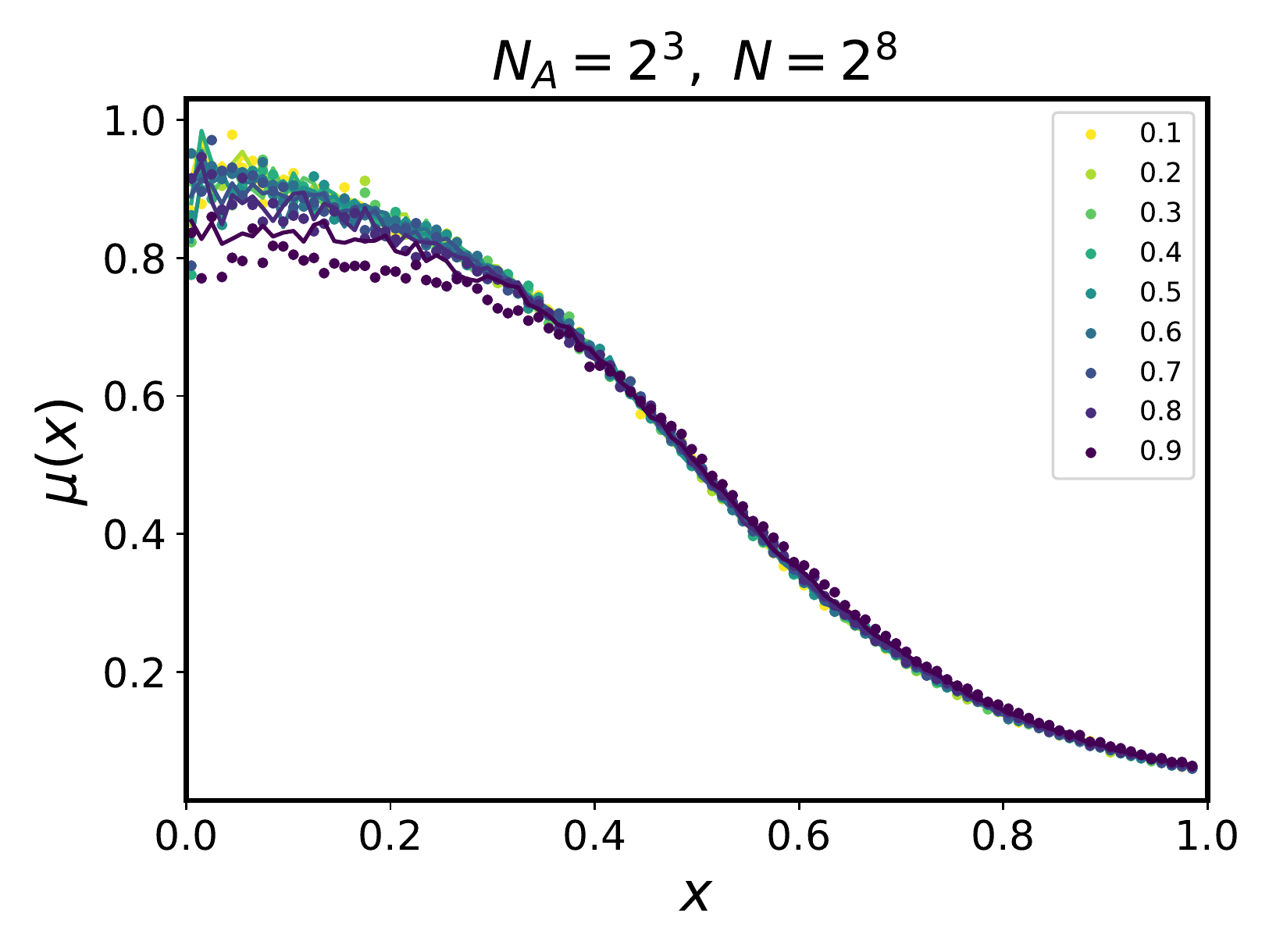}
  \includegraphics[width = .48\textwidth]{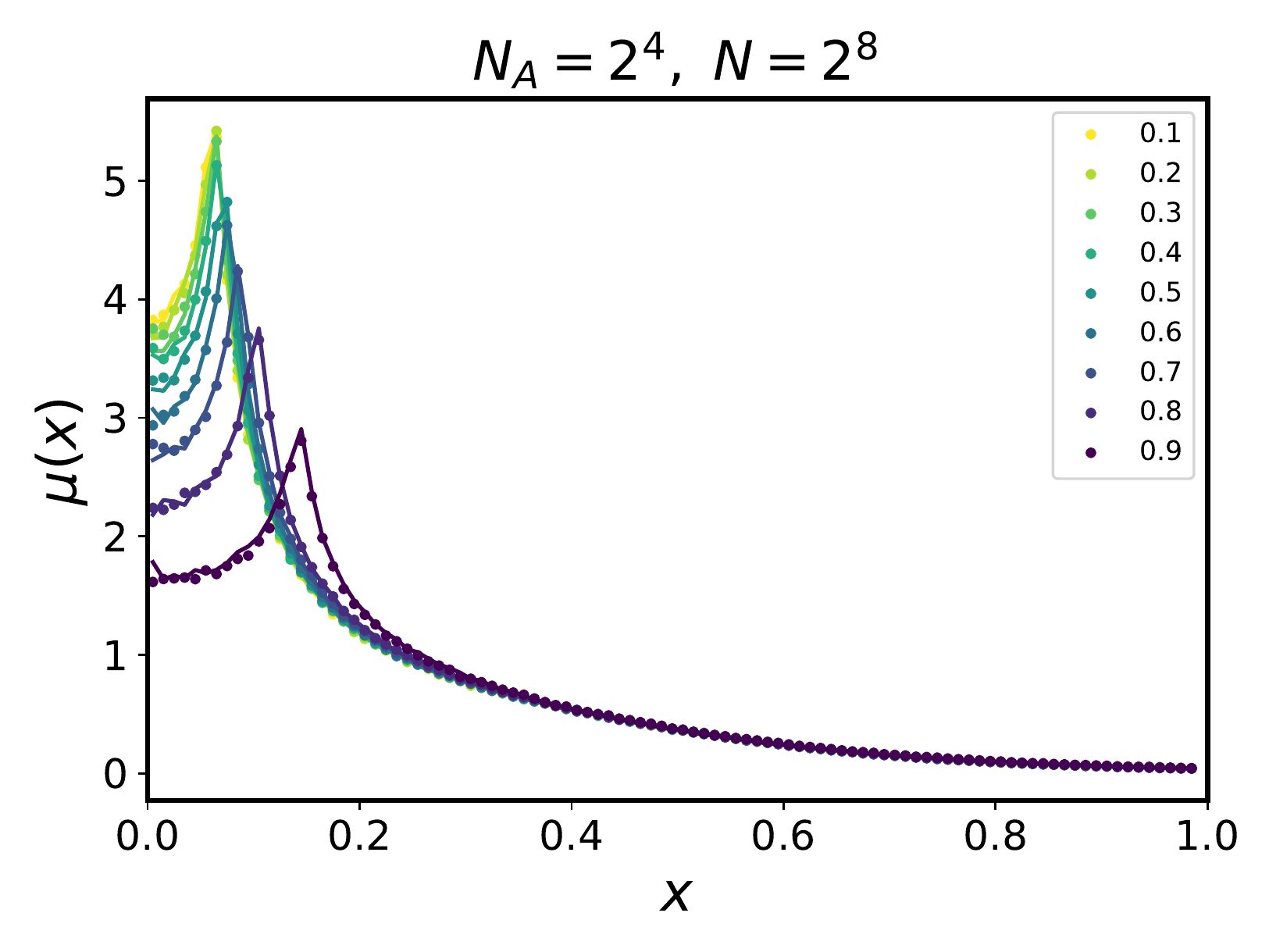}
  \caption{The spectrum of equation (16) of the main text computed numerically (solid lines) as compared to numerical data (dots) for a range of $|z|$. }
  \label{spectrum_N8_full_fig}
\end{figure}

In the main text, we focused on the parameter regime that is of the most physical relevance, that of the thermodynamic limit ($N_A , N_B \rightarrow \infty$) and not perfectly equal subsystems ($N_A \neq N_B$). In this limit, the structure of the reduced density matrix simplified to a randomly rescaled and deterministically shifted Ginibre matrix. This simplification allowed us to evaluate the spectrum analytically. In the main text, we demonstrated that this asymptotic answer agreed well with numerical data from small matrices in the regime of interest. However, there were clear deviations when $N_A$ became small or close to $N_B$. In Figure \ref{spectrum_N8_full_fig}, we show that these deviations are precisely accounted for by considering the full structure of the density matrix given in equation (16) of the main text, valid in all parameter regimes away from $|z| \sim 1$. When $N_A = N_B$, we find that there is strong eigenvalue dependence in the spectrum.

\section{On the definition of von Neumann entropy}

In the main text, we used a non-standard definition of von Neumann entropy, suitable for non-Hermitian systems involving taking an absolute value. Here, we demonstrate the answer for the standard definition of von Neumann entropy. This is straightforward because we have already solved for the complete eigenvalue spectrum of $\rho_A$. If we choose to take the usual definition of the entanglement entropy, the argument of the logarithm is complex and there is an ambiguity in which branch we take. We have
\begin{align}
  \E{M}{-N_A \int d\lambda \mu(\lambda)\lambda \log \lambda} = -\frac{2N_A}{\pi R^2}\int_{-\pi}^{\pi} d\theta\int_0^R dr {r} \left(N_A^{-1}+ r e^{i\theta} \right)
  \log \left(N_A^{-1}+ r e^{i\theta}\right).
\end{align}
The principal value of the logarithm is
\begin{align}
    \mathcal{PV} \left[\log \left(N_A^{-1}+ r e^{i\theta}\right)\right] = \frac{1}{2}\log \left(N_A^{-2}+\frac{2 r \cos (\theta )}{N_A}+r^2\right)+i \sin ^{-1}\left(\frac{r \sin (\theta )}{\sqrt{N_A^{-2}+N_A^{-1}{2 r \cos
   (\theta )}+r^2}}\right).
\end{align}
Integrating the first term gives the entropy we studied in the main text. The second term is odd under $\theta \rightarrow -\theta$. This means that the imaginary piece of the integral will vanish because the real part of $N_A^{-1}+re^{i\theta}$ is even under $\theta \rightarrow -\theta$. The imaginary part of $N_A^{-1}+re^{i\theta}$ is odd, so we must evaluate
\begin{align}
     \frac{2N_A}{\pi R^2}\int_{-\pi}^{\pi} d\theta\int_0^R dr {r^2} \sin \theta 
 \sin ^{-1}\left(\frac{r \sin \theta }{\sqrt{N_A^{-2}+N_A^{-1}{2 r \cos
   \theta }+r^2}}\right) = \frac{8N_A R}{3\pi},
\end{align}
where we have only included the leading piece in $N_A$.
Integrating over $\gamma_2$
\begin{align}
    \int_0^{\infty} d\gamma_2 e^{-\gamma_2}\gamma_2^{1/2} \frac{8N_A \sqrt{1-|z|^2}}{3\pi }
    = \frac{4N_A\sqrt{1-|z|^2}}{3\sqrt{\pi}}
\end{align}
Therefore, the entropy is unnaturally large, scaling exponentially with the number of degrees of freedom.

\end{document}